\documentclass[10pt,journal]{IEEEtran}

\usepackage[nocompress]{cite}
\usepackage{algorithm2e}
\usepackage{tikz}
\usepackage{pifont}
\usepackage{pgfplots}
\usepackage{xspace}
\usepackage{caption}
\usepackage{subcaption}
\usepackage{xcolor,colortbl}
\usepackage{booktabs}
\usepackage{url}
\usepackage{amsmath}
\usepackage{paralist}
\newcommand{\ours}{\textit{InterSnap}\xspace}
\usepackage{enumitem}
\usepackage{mathtools}
\usepackage{amssymb}
\usepackage{graphicx}   
\usepackage{xcolor}     
\usepackage{transparent}

\newcommand{\notets}[1]{{\color{black}#1}}

\begin{document}


	%

\title{Auditable Ledger Snapshot for Non-Repudiable Cross-Blockchain Communication}

\author{Tirthankar Sengupta,~\IEEEmembership{Student~Member,~IEEE},~Bishakh~Chandra~Ghosh,~\IEEEmembership{Member,~IEEE},~Sandip~Chakraborty,~\IEEEmembership{Senior~Member,~IEEE,} and~Shamik~Sural,~\IEEEmembership{Senior~Member,~IEEE}

 \thanks{Tirthankar Sengupta, Sandip Chakraborty, and Shamik Sural are with the Department of Computer Science and
Engineering, Indian Institute of Technology, Kharagpur, India (mailtisen03@kgpian.iitkgp.ac.in, sandipc@cse.iitkgp.ac.in, shamik@cse.iitkgp.ac.in). Bishakh Chandra Ghosh is with Pinggy Technology Private Limited, India (ghoshbishakh@gmail.com). }
\thanks{This work has been supported by Anusandhan National Research Foundation (ANRF), previously Science and Engineering Research Board (SERB), Government of India, Core Research Grant through grant number CRG/2022/002565 dated 17 January 2023.}
}


\markboth{IEEE Transactions on Services Computing,~Vol.~XX, No.~Z, Month~YYYY}%
{Sengupta \MakeLowercase{\textit{et al.}}: InterSnap: A Secure Non-Repudiable Snapshot
Transfer for Interoperable Permissioned Blockchains}

 \IEEEtitleabstractindextext{
	\begin{abstract}
Blockchain interoperability is increasingly recognized as the centerpiece for robust interactions among decentralized services. Blockchain ledgers are generally tamper-proof and thus enforce non-repudiation for transactions recorded within the same network. However, such a guarantee does not hold for cross-blockchain transactions. When disruptions occur due to malicious activities or system failures within one blockchain network, foreign networks can take advantage by denying legitimate claims or mounting fraudulent liabilities against the defenseless network. In response, this paper introduces \ours, a novel blockchain snapshot archival methodology, for enabling auditability of cross-blockchain transactions, enforcing non-repudiation. \ours introduces cross-chain transaction receipts that ensure their irrefutability. Snapshots of ledger data along with these receipts are utilized as non-repudiable proof of bilateral agreements among different networks. \ours enhances system resilience through a distributed snapshot generation process, need-based snapshot scheduling process, and archival storage and sharing via decentralized platforms. Through a prototype implementation based on Hyperledger Fabric, we conducted experiments using on-premise machines, AWS public cloud instances, as well as a private cloud infrastructure. We establish that \ours can recover from malicious attacks while preserving cross-chain transaction receipts. Additionally, our proposed solution demonstrates adaptability to increasing loads while securely transferring snapshot archives with minimal overhead.





\end{abstract}}

	\maketitle

	\IEEEdisplaynontitleabstractindextext

	%
	\IEEEpeerreviewmaketitle


\section{Introduction}
\label{sec:intro}
\IEEEPARstart{I}{nter}-blockchain communication \cite{ren2023interoperability} is emerging as a vital feature for almost all decentralized applications (dApps) in recent times. For instance, \textit{Uniswap}~\cite{adams2021uniswap} and \textit{Binance}~\cite{binance2022} 
leverage cross-chain interoperability to facilitate exchange of tokens and digital assets across different blockchain networks without relying on any centralized authority. Inter-blockchain ecosystems such as \textit{Polkadot}~\cite{wood2016polkadot} and \textit{Cosmos}~\cite{kwon2019cosmos} have been providing a robust infrastructure for handshaking of messages and assets between disparate blockchains. 
Industry-grade interoperable solutions, such as IBM Food Trust \cite{ibmfoodtrust}, leverage blockchain technology to provide a verifiable and immutable record of food production, processing, and distribution of data across the supply chain.
Major food companies such as Walmart, Nestle, and Dole are using it for tracking their procurement and distribution network.
Several existing works \cite{margheri2020decentralised, akbarfam2024sok, pan2023data,ramachandran2018smartprovenance,zhang2022poster} discuss cross-chain provenance in collaborating systems.
Many blockchain interoperability frameworks have been developed over the years \cite{belchior2021survey, ietf-satp-core-07}. In the permissioned blockchain space, projects such as Hyperledger Cacti~\cite{cactis} and Secure Asset Transfer Protocol (SATP) \cite{ietf-satp-core-07} facilitate communication and data exchange between different blockchain networks. Collectively, these instances underscore a growing interest in inter-blockchain communication-based services in recent times.


In cross-chain environments involving both public and permissioned blockchains, a fundamental asymmetry exists in verifiability. When a transaction originates from a public blockchain, a permissioned blockchain can directly validate it by inspecting the publicly accessible state(s)~\cite{li2025towards}. In contrast, when a permissioned blockchain initiates a transaction, the public blockchain cannot independently verify it~\cite{abebe2021verifiable}, as access to the private ledger is inherently restricted. This limitation becomes even more pronounced in private-to-private blockchain interoperability. For example, in cross-chain transactions between a Hyperledger Fabric network~\cite{hlf} and a Corda network~\cite{brown2016corda} or even between two Hyperledger Fabric networks, neither network can access the ledger data or transaction history of the other, since both ledgers enforce strict access controls, and issues of accountability and false claims arise because the participants of one network do not have direct visibility into the other network.

These limitations in verifiability and the lack of direct visibility between networks create opportunities for malicious behavior. Even if participants behave honestly within their own network, a sufficient number of them may still collude to attack a foreign network with which they are interoperating. Such coordinated misconduct is plausible, and even likely, when participants prioritize self-interest over fairness toward other networks. In such cases, a malicious network can unjustly deny legitimate claims from interconnected networks and disregard prior transactions or commitments. The risk is further amplified when a catastrophic failure of a blockchain system~\cite{failurecoinbase, failuresolana} causes partial loss of its inter-network transaction history. Without these records, other interoperating networks may exploit the situation by making wrongful payment demands, engaging in double spending, or issuing false claims, knowing there is no evidence to disprove them.


Addressing the interoperability challenges requires implementing a redundancy mechanism that preserves a non-repudiable ledger of cross-chain transactions. Consider an example scenario of two interoperating blockchain networks, $\mathcal{N}_1$ and $\mathcal{N}_2$. A critical situation occurs when $\mathcal{N}_2$ encounters a sudden system-wide failure, causing a partial loss of historical ledger data. Exploiting the vulnerability presented by $\mathcal{N}_2$'s system failure, most participants of $\mathcal{N}_1$ take a strategic and potentially malicious approach. One of the participants of $\mathcal{N}_1$, say Party A, seizes this opportunity to add fraudulent payments against Party B of $\mathcal{N}_2$, gets it endorsed by other authorized peers, and commits them to the ledger of $\mathcal{N}_1$. $\mathcal{N}_1$ then disseminates them to $\mathcal{N}_2$ through cross-chain transactions and demands the return of payments or assets from Party B. So, in this case, disruptive behavior has two roots - a) $\mathcal{N}_2$ suffers a failure, causing loss of data, and b) One or more participants of $\mathcal{N}_1$ have malicious intention against $\mathcal{N}_2$. Without any cross-chain transaction history available to verify Party A's demand, Party B succumbs to the false claim by $\mathcal{N}_1$. Such motivation behind $\mathcal{N}_1$'s demand could range from seeking financial gain to defaming $\mathcal{N}_2$. This scenario illustrates only one of the intricate dynamics of inter-blockchain communication and the potential risks associated with failures and malicious faults.

Data interoperability solutions, such as SATP~\cite{ietf-satp-core-07}, rely on gateways and use receipts with two-phase commit protocols to ensure reliable cross-chain data transfer. Similarly, existing snapshot mechanisms, including Fabric Snapshots~\cite{hlfledgersnapshot} and BUNGEE~\cite{belchior2024bungee}, can help mitigate ledger data loss. However, these approaches do not record proofs of historical cross-chain transactions, which are essential for enforcing accountability. As a result, current systems remain vulnerable to fraudulent claims and cannot guarantee non-repudiation due to the following key limitations.

{
\setlength{\leftmargini}{0cm}
\begin{compactenum}[i)]
\item \textbf{No inter-blockchain history:} Typical blockchain transactions infallibly become part of the ledger. Contrarily, inter-blockchain assets, and data transfer details are not always recorded. Proofs in the forms of signatures and attestations of foreign networks' participants are crucial for the auditability of these cross-chain transactions.

\item \textbf
{Lack of snapshot transfer mechanism:} Existing systems do not have any means of sharing snapshot data with other networks. The lack of an inter-blockchain snapshot transfer mechanism prevents external auditors from investigating a dispute. The transfer of snapshots is crucial for collaborative efforts to prevent inter-blockchain transaction fraud.
\item \textbf
{Limited snapshot records:}
Typical snapshots capture only the current state of the blockchain. Even if the block headers and the current state are archived, the entire transaction data is often pruned, making it almost impossible to refute fraudulent claims by other networks.

\item \textbf{Consistency and recency issues:}
Existing systems often employ an ad-hoc node to record the snapshots \cite{hlfledgersnapshot}. This results in a high risk of a fork, leading to inconsistency between the current state of the blockchain and historical data. Moreover, long intervals between snapshots often cause a loss of recent transaction data in case of a failure.
\end{compactenum}
}

\noindent Resolving the above limitations, in this paper, we develop \ours, a non-repudiable cross-network archive exchange architecture for fault tolerance in permissioned blockchains. The proposed framework builds on a preliminary version of a snapshot transmission method we introduced in our earlier work \cite{sengupta2023cross}. 
In the present context, we introduce non-repudiable receipts for cross-chain transactions. \ours includes these receipts in snapshots to facilitate transparency and resolve disputes between disparate blockchain networks. We also improve the existing snapshot and storage mechanisms for a more reliable archival of ledger data. Our contributions are highlighted below.
{
\setlength{\leftmargini}{0.5cm}
\begin{compactenum}

 \item \textbf{Inter-blockchain transaction archival.} \ours proposes the first protocol for cross-blockchain transaction archival. For every cross-network message (data/asset) initiated by a sender, \ours enforces collection of a transaction receipt endorsed by the receiver network. Each cross-chain transaction, along with the corresponding receipt, endorsements, and attestations, are recorded in snapshots as proof to enforce non-repudiation.
    \item \textbf{Inter-blockchain dispute resolution.} \ours introduces an independent auditor-based resolution mechanism for disputes arising between interoperating blockchain networks. The auditor here is a separate blockchain network comprising reputed organizations, such as audit firms, who use the snapshots from both the disputing networks to decide a case.
    \item \textbf{Snapshot peer selection:} 
    Contrary to the existing norm of selecting an arbitrary peer for the snapshot procedure, we propose a distributed peer selection process based on their readiness and block height. This ensures faster snapshots while capturing the oldest available transaction data.  
    
    \item \textbf{Need-based snapshot scheduling:} It is crucial to determine when to take a snapshot. Instead of just relying on periodic snapshots, in this paper, we develop need-based scheduling that identifies significant changes in the ledger to take a snapshot. This optimizes both snapshot latency and archive size. 

    \item \textbf{Distributed archive storage and sharing:} 
    We design a mechanism for encrypting and sharing the snapshot archives over the decentralized storage platform - Interplanetary File Systems (IPFS) \cite{ipfsdc} aligned with Web 3.0 \cite{web3storage}. This enables inter-network snapshot transfer for external auditors, facilitating dispute resolution.
 
\end{compactenum}
}
Through our implementation using Hyperledger Fabric~\cite{hlf} and Hyperledger Cacti~\cite{cactis}, we demonstrate that \ours can adapt to increasing loads and transfer snapshot archives from the source to the destination network with modest resource usage. Although \ours uses established techniques such as IPFS, Service Discovery Protocol, and Hyperledger Cacti, the true strength of the robust cross-chain defense system emerges when these components are amalgamated with snapshots to provide an optimized, cross-chain defense against attacks and failures across blockchain networks.


\begin{table*}[!t]
\centering
\tiny
\caption{Comparative Analysis}
\label{table:crosschain_comparison}
\begin{tabular}{p{2.6cm}p{4.3cm}p{1cm}p{1.6cm}p{1.5cm}p{1.5cm}p{1.2 cm}p{1 cm}}
\toprule
  \textbf{Research Contributions} & \textbf{Core Mechanism} & \textbf{Distributed Peer Selection} & \textbf{Periodic Schedule} & \textbf{Cross-Blockchain Snapshot Exchange} & \textbf{Blockchain Type} & \textbf{Cross-chain Acknowledgment} & \textbf{Receipt Transaction} \\
  \midrule
   Marsalek et al. \cite{marsalek2021compressing} & UTXO-based Snapshot Chain & No & Yes & No & Public & No & No \\
   Luo et al. \cite{luo2024crosschannel} & Cross-org payment info storage with snapshots & No & No & No & Permissioned & Yes & No \\
   Abebe et al. \cite{abebe2021verifiable} & Verifiable cross-chain observation & No & No & Yes & Permissioned & Yes & No \\
   Belchior et al. \cite{belchior2024bungee} & Blockchain view system & No & Yes & Yes & Permissioned & Yes & No \\
   Hardjono et al. \cite{ietf-satp-core-07} & Asset Transfer Protocol for blockchain gateways (SATP) & No & No & No & Both & Yes & No \\
   Kwon et al. \cite{kwon2019cosmos} & Cosmos IBC & No & No & No & Public & Yes & No \\
   Breidenbach et al. \cite{breidenbach2021chainlink} & Chainlink CCIP & No & No & No & Public & Yes & No \\
   Wood et al. \cite{wood2016polkadot} & Polkadot XCMP & No & No & No & Public & Yes & No \\
   Herlihy et al. \cite{herlihy2018atomic} & Inter-blockchain asset exchange & No & No & No & Public & Yes & No \\
   Han et al. \cite{han2023vm} & Cross-chain smart contract & No & No & No & Both & No & No \\
   Han et al. \cite{han2021vassago} & Inter-blockchain provenance query & No & No & No & Permissioned & Yes & No \\
   Abebe et al. \cite{abebe2019enabling} & Inter-blockchain Connector & No & No & No & Permissioned & Yes & No \\
   \rowcolor{blue!30}\textbf{\ours} & {\raggedright Inter-blockchain Snapshot Exchange} & \textbf{Yes} & \textbf{Yes} & \textbf{Yes} & \textbf{Permissioned} & \textbf{Yes} & \textbf{Yes} \\
\bottomrule
\end{tabular}
\end{table*}

The rest of the paper is structured as follows. In Section \ref{sec:relatedwork}, we review the existing literature. Section \ref{sec:faultScenarioRecovery} presents the fault model and trust model. \ours architecture and methodology are detailed in Section \ref{solutionapproach_architecture}. The security analysis of \ours are discussed in Section \ref{sec:analysis}, followed by details of its implementation in Section \ref{sec:implemenation}. The results of our experiments are presented in Section \ref{sec:results}, before concluding with future directions in Section \ref{sec:conclusion_futurework}.

\section{Related Work}
\label{sec:relatedwork}
Snapshots play a crucial part in blockchain systems, particularly for state preservation and recovery.  They are widely used in blockchains as a low-overhead solution \cite{hlfledgersnapshot,gorenflo2020fastfabric,ali2018blockchain,krug2015sidecoin} for initializing and synchronizing new nodes.
We review existing literature on snapshot storage and potential inter-blockchain transfer approaches.

\smallskip

\noindent \notets{\textbf{Snapshot Formation Approaches:}}
Sun et al. \cite{sun2021sasledger} proposed a snapshot storage solution utilizing a remote server for preserving copies of snapshots. On the other hand, the work in~\cite{pi2021xfabledger} advocated for storing ledger state content to a central server while maintaining a commit reference in the peer.
Ren et al. \cite{ren2021snapshotsave} proposed a snapshot synchronization method based on  UTXO. The existing UTXOs, at a specific block height, are transferred to a new peer to rebuild the system state. 
Marsalek et al.
\cite{marsalek2021compressing} proposed a periodic creation of a snapshot chain connected to the main chain to store compressed state periodically using UTXO for the bitcoin blockchain. 

\smallskip

\noindent \notets{\textbf{State Sharing and Views:}}
Abebe et al.~\cite{abebe2021verifiable} proposed a state-sharing approach for permissioned networks operating under adversarial conditions, where a majority of committee members may be malicious. Their method employs a public-ledger-based bulletin board and trusted observer to mitigate such behavior. More recently, Belchior et al.~\cite{belchior2024bungee} introduced a global blockchain view system, BUNGEE, which provides stakeholder-specific ledger snapshots. This approach offers standardized view formats to promote interoperability, while allowing sensitive data to be obfuscated during snapshot creation. BUNGEE can serve as a pluggable component in \ours for generating snapshots. However, on their own, these approaches cannot fully address the challenge of ensuring non-repudiation in bilateral contracts after ledger data loss.

\smallskip

\noindent \notets{\textbf{Blockchain Data Storage Approaches:}}
A blockchain can store the complete transaction history in the ledgers of both leader and subordinate peers~\cite{raft}, an approach commonly referred to as the \textbf{on-chain} method. This can also be combined with sharding, where the blockchain network is divided into smaller peer groups or sub-units~\cite{zamani2018rapidchain, caismartchain, set2022service, xu2023two}. Sharding reduces the need to store the entire dataset on every node and improves parallelization, but each node within a shard must still maintain a full copy of its shard’s blocks~\cite{wang2019monoxide, chen2019sschain}, adding storage overhead. Alternatively, Consensus Unit (CU)-based solutions~\cite{xu2018cub} have been proposed, where each node stores different blocks~\cite{xu2018cub, dai2019jidar}, unlike shard nodes that store identical shard data. However, these storage methods are generally more suitable for preserving only a limited set of interlinked transactions or data. In the \textbf{off-chain} approach, part of the data is stored outside the main blockchain. Gorenflo et al.~\cite{gorenflo2020fastfabric} proposed using a distributed database to optimize storage and validation while reducing I/O overhead. Ali et al.~\cite{ali2018blockchain} improved efficiency with a DHT-based distributed ledger, while Pi et al.~\cite{pi2021xfabledger} used a central remote database for local peer data. IPFS-based solutions have also emerged: Azbeg et al.~\cite{azbeg2022blockmedcare} combined IPFS with a public blockchain for secure healthcare, Kumar et al.~\cite{kumar2019implementation} employed content addressing in IPFS to enhance tamper-resistance, and Zhang et al.~\cite{zhang2022poster} applied signcryption to improve public verifiability.

\smallskip

\noindent \notets{\textbf{Interoperability Frameworks:}}
Interoperability approaches have also been significantly explored in recent cross-chain research endeavors. 
In atomic cross-chain swaps (also labeled as \textit{atomic swaps})
~\cite{herlihy2018atomic, ren2023interoperability},
certain amount of assets is destroyed on the source blockchain, and the same amount is re-created on the destination blockchain as an atomic process \cite{belchior2022hermes}.
However, the long wait needed to release the timelocks 
may pose an issue due to the rapid volatility in case of cryptocurrency values.
To overcome this, Jin et al. elaborate a different blockchain interoperation scheme \cite{jin2018towards} consisting of an active and a passive mode. 
Han et al. \cite{han2023vm} proposed a cross-chain smart contract execution scheme with the help of migrating virtual machines (VMs) from one blockchain to another blockchain using containers. 
Inter-blockchain provenance query architecture
\cite{han2021vassago} is also proposed, but it is prone to a single point of failure due to the reliance on shared blockchain nodes.
In the Side-chain approach \cite{back2014enabling}, a secondary blockchain runs independently of the main blockchain. 
Through cross-chain communication, the secondary chain can actively read and verify information, perform actions \cite{yin2023survey}, and then synchronize with the main chain. 
In the Notary scheme \cite{scheid2021notary,koens2019notary}
to conduct a cross-chain transaction, related parties can submit transactions to their respective blockchain-trusted notary nodes for sign-off. Alternatively,
the Relay-based techniques
do not depend solely on the trusted third party and only collect the data status of different chains through intermediaries for self-verification \cite{frauenthaler2020leveraging}. 
Interoperability solutions such as Cosmos \cite{kwon2019cosmos} and the relay-chain in Polkadot blockchain Para-chain \cite{wood2016polkadot} use relay-based systems.
On the other hand, the Blockchain Connector ~\cite{abebe2019enabling} category comprises interoperability solutions that are not cryptocurrency-directed or blockchain engine-based. 
They utilize trusted relays to discover the target blockchain communications
This approach has been followed in the IBM DLT Weaver framework \cite{dltinterop}, which is merged in Hyperledger Cacti \cite{cactis}. 
Cacti is an open-source interoperability framework designed to enable secure, cross-chain asset and data exchange between heterogeneous blockchain networks. It supports a modular architecture that uses drivers, relay services, and interoperable chaincode to connect networks such as Hyperledger Fabric, Corda, and other external systems.
On the other hand, SATP \cite{ietf-satp-core-07} protocol also provides a framework for inter-blockchain data transfer. 
It employs receipts for two-phase commits between two gateways for initiating data/asset exchange.
However, cross-blockchain transaction non-repudiation has not been addressed in the existing research. 
Table \ref{table:crosschain_comparison} compares \ours with existing approaches. While some cross-chain methods support snapshot and data exchange, none integrate receipt transactions, distributed peer selection, periodic scheduling, and cross-chain snapshot exchange. This makes \ours a comprehensive solution for secure, non-repudiable cross-chain communication.

\section{System Model and Challenges}
\label{sec:faultScenarioRecovery}
We consider the inter-operation between any pair of different permissioned blockchain networks, say, $\mathcal{N}_1$ and $\mathcal{N}_2$.
Data originating from a network $\mathcal{N}_1$ and transferred to $\mathcal{N}_2$ is trusted and agreed upon by the majority of peers of $\mathcal{N}_1$. This is because this data passes through the consensus rules of $\mathcal{N}_1$, and thus constitutes data already committed in its ledger.
However, the recipient network $\mathcal{N}_2$ does not implicitly trust the external transactions coming from the foreign network $\mathcal{N}_1$. 
Indeed, \ours accounts for both failures and malicious behavior at the level of entire networks. By malicious behavior, we refer to situations where a network collectively acts against another network. While blockchain systems are generally resilient to internal adversarial actions because of consensus protocols that defend against crash and Byzantine faults, nodes within a network may still coordinate to act with malicious intent toward an interoperating network. In this case, the internal nodes are not adversaries to their own network, but rather the network as a whole becomes an adversary to another. For example, if $\mathcal{N}_1$ crashes and loses data, $\mathcal{N}_2$ may exploit the situation by maliciously denying a legitimate contractual commitment made prior to the crash.


We presume that the failure of participant nodes in the blockchain networks does not imply the failure of the IPFS network because it is a separate system with its own redundant nodes. Therefore, existing snapshot archives stored in the IPFS remain intact regardless of the status of the blockchain nodes. 
We denote a cross-chain transaction as $T_{cc}$, and a corresponding response transaction as $T_{rt}$, together forming a cross-chain transaction set
$\mathcal{T}_{\text{Set}} = \{T_{cc}, T_{rt}\}
$.
After $T_{cc}$ is initiated by a peer, say, ${p}_i$ in $\mathcal{N}_1$, the transaction is sent for endorsement. The collection of these endorsements is represented as 
$
\psi_{\mathcal{N}_1}(T_{cc}) = \bigcup_{i=1}^{k} \text{Endorsements}_{p_i}(T_{cc})
$, where $k$ represents the total number of endorsing participants.
If the number of endorsements satisfies the majority rule, such that $|\psi_{\mathcal{N}_1}(T_{cc})| \geq \frac{2|\mathcal{N}_1|}{3}$, where $|\mathcal{N}_1|$ is the total number of peers in $\mathcal{N}_1$, the transaction should be deemed valid and committed to the ledger $\mathcal{L}_{\mathcal{N}_1}$. The height of the ${\mathcal{N}_1}$ ledger is denoted as $|\mathcal{L}_{\mathcal{N}_1}|$. Once committed, the transaction payload along with endorsements, $T_{cc}||\psi_{\mathcal{N}_1}(T_{cc})$ is dispatched for cross-chain propagation. 
Upon receiving $T_{cc}$, the participants in $\mathcal{N}_2$, first verify the endorsements $\psi_{\mathcal{N}_1}(T_{cc})$ to confirm that the transaction was agreed by $\mathcal{N}_1$. If the majority of participants in $\mathcal{N}_2$ can accept $T_{cc}$, the collection of endorsements of $l$ endorsing peers of $\mathcal{N}_2$ can be represented as $
\psi_{\mathcal{N}_2}(T_{cc}) = \bigcup_{j=1}^{l} \text{Endorsements}_{p_j}(T_{cc} \| \psi_{\mathcal{N}_1}(T_{cc}))
$, where $p_{j}$ is a peer node in $\mathcal{N}_2$.
The transaction can be committed to the ledger $\mathcal{L}_{\mathcal{N}_2}$ and generate a response transaction $T_{rt}$ containing the signed response from the peers of $\mathcal{N}_2$. 
So, the endorsed response $T_{rt}||\psi_{\mathcal{N}_2}(T_{cc})$
is to be sent back to  $\mathcal{N}_1$ as an acknowledgment. If both valid $T_{cc}$ and $T_{rt}$ are present,
$\mathcal{T}_{\text{Set}}$ can be marked as complete.
Let $S_{\mathcal{N}_1}^t$ represent the snapshot of the ledger in network $\mathcal{N}_1$ at time $t$. After the cross-chain transaction $T_{cc}$ and $T_{rt}$ are committed, the next snapshot, $S_{\mathcal{N}_1}^{(t+1)}= \{T_{cc}, T_{rt}\}$. So, the snapshot archive $S_{arc}$ at timestamp $t+m$ is thus updated as 
$S_{\mathcal{N}_1}^{t} \cup S_{\mathcal{N}_1}^{(t+1)} \dots S_{\mathcal{N}_1}^{(t+m)} $.

We assume the auditor is a trusted third-party organization, similar to real-world independent institutions such as certified audit firms, dispute resolution boards, or governance tribunals, which are typically responsible for mediating financial and compliance disputes.
\notets{The auditor has access to snapshots from all participating networks and is assumed to provide honest verification during disputes.
}
Next, we present the trust model and the fault model considered while designing \ours. 

\subsection{Trust Model}
\notets{\ours assumes that intra-network consensus enforces trust within the consortium. However, this trust does not automatically extend beyond network boundaries, as one network does not have any direct view to the data or consensus state of the other network.}

\subsubsection{\textbf{Intra-Network Trust Model}}
Within a network, the consensus model is based on endorsements \cite{androulaki2018hyperledger}. Even if some participants of the network are faulty, as long as more than $\frac{2}{3}\mathsf{rd}$ of the peers within the network are non-faulty, the system can endure Byzantine attacks \cite{lamport1982byzantine, castro1999practical}.
If a transaction receives endorsements from a majority of peers within the network, it is deemed trustworthy and can be validated as legitimate. This ensures the integrity of internal transactions and prevents fraudulent activities within the network. So, for a cross-chain transaction $T_{cc} $ initiated by a participant of $\mathcal{N}_1$, its peers first agree upon the transaction invoked from the $\mathcal{N}_1$ side through its own consensus process. The said transaction is endorsed and committed in the ledgers of $\mathcal{N}_1$. We use majority endorsement $|\psi_{\mathcal{N}_1}(T_{cc})|$ on the smart contract level. Next, the $T_{cc}$ is disseminated to $\mathcal{N}_2$. 
$\mathcal{N}_2$, upon receiving the transaction, validates it by its peers, and if the majority agrees, then that transaction is accepted and committed.

\subsubsection{\textbf{Inter-Network Trust Model}}\label{subsec:trustmodel}
For inter-network transactions involving multiple blockchain networks, the networks do not have any direct view of the consensus of the foreign networks. When $\mathcal{N}_2$ receives cross-chain transactions from $\mathcal{N}_1$, it has to rely on the proofs and attestations in the form of digital signatures to validate if the data has been endorsed by the majority of $\mathcal{N}_1$'s participants or not. Even then it cannot automatically deem the content as absolute truth. This caution arises from the possibility that
although $\mathcal{N}_1$ is resistant to Byzantine behavior internally, the majority of peers of $\mathcal{N}_1$, collectively might have malicious intent towards $\mathcal{N}_2$. Such a malicious stance of the entire network towards a foreign network can potentially result in the transmission of altered information. In such a scenario, $\mathcal{N}_2$ must exercise additional vigilance to refute any fraudulent claims by $\mathcal{N}_1$.

\subsection{Fault Model}
\label{subsec:faultmodel}
In general, we consider that some peers in a network of $n$ nodes may face faults. However, we assume that the number of Byzantine faulty nodes at an instance never exceeds  $\frac{n}{3} - 1$. The following fault scenarios are taken into account while designing \ours.
\subsubsection{\textbf{Malicious Faults Resulting Breach of Contract}}
In inter-blockchain transactions, there exists a risk of malicious behavior by one network, leading to contract violations. Consider a situation when a participant from $\mathcal{N}_1$ has provided a service or asset transfer to a participant in $\mathcal{N}_2$. The latter has contractual obligations for the return of assets or services as stipulated in the contract. It is possible that malicious actors enjoying the majority within $\mathcal{N}_2$ can deny $T_{cc}$ by not fulfilling their contractual agreement towards $\mathcal{N}_1$. This denial could be intentional and may occur by refusing to acknowledge the original transactions and assets, delaying the processing of the payment, etc.
\subsubsection{\textbf{Loss of Ledger Data Causing Unintended Violation of Contract}}
Data loss in one blockchain network can cause unintentional violation of contracts with other interoperating networks. Consider $\mathcal{N}_2$ faces a catastrophic failure.
This failure could stem from technical glitches, network outages, or other unforeseen circumstances, disrupting the normal flow of operations within $\mathcal{N}_2$. As a result, parts of the ledger data are lost in  $\mathcal{N}_2$. After the revival of $\mathcal{N}_2$ and restoration of interconnection, for a certain transaction $T_{cc}$, if a participant of $\mathcal{N}_1$ requests a participant of $\mathcal{N}_2$ to fulfill some past liability, $\mathcal{N}_2$ fails to comply, as it has no pre-existing reference to its past history. This situation potentially causes financial losses or disputes between the involved consortia. 
\subsubsection{\textbf{Fraudulent Exploitation of System-wide Failure}}
Suppose $\mathcal{N}_2$ suffers from catastrophic failure and resurrects afresh with a partial or total loss of historical data.  
Recognizing this vulnerability, the majority of peers of $\mathcal{N}_1$ seize the opportunity to maliciously take advantage of  $\mathcal{N}_2$'s failure.  $\mathcal{N}_1$ may generate counterfeit transactions, which are then used to demand fraudulent dues from $\mathcal{N}_2$. This places $\mathcal{N}_2$ in a precarious situation, as it must confront unwarranted financial obligations.
\par

\section{InterSnap Architecture}
\label{solutionapproach_architecture}
Snapshots are used in permissioned blockchains to efficiently bootstrap new peers joining a network \cite{hlfledgersnapshot, hlf}. However, substantial improvements and modifications are required over such existing snapshot techniques in order to utilize the snapshots for defending against malicious inter-blockchain attacks. \ours addresses these requirements through the following six pillars:
(A) Recording Non-Repudiable Transaction Receipts, 
(B) Need-based Snapshot Scheduling, 
(C) Selection of Snapshot Generating Peer,  (D) Archive Encryption, (E) Saving Encrypted Archives in Decentralized Storage,
and (F) Data Interoperability.

\begin{figure}[!t]
\centering
\centering
{\includegraphics[width=0.8\linewidth]{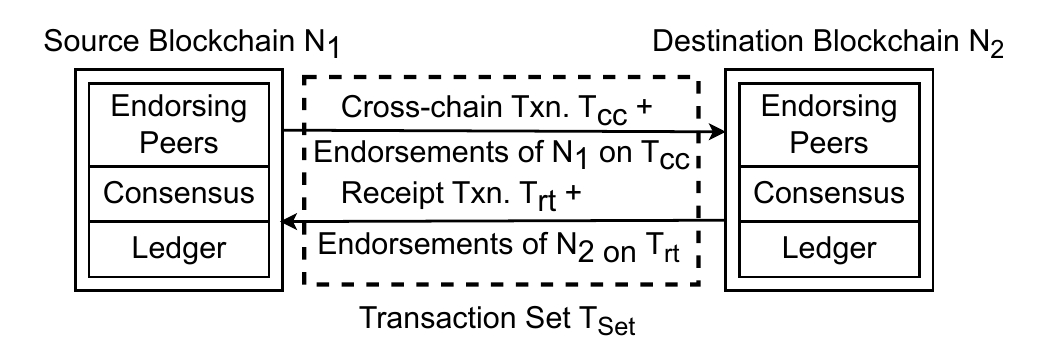}}
\caption{Cross-Chain Transaction Receipts}
\label{fig:crosschaintransaction}
\end{figure}

\begin{figure}[!ht]
\centering
{\includegraphics[width=1\columnwidth]{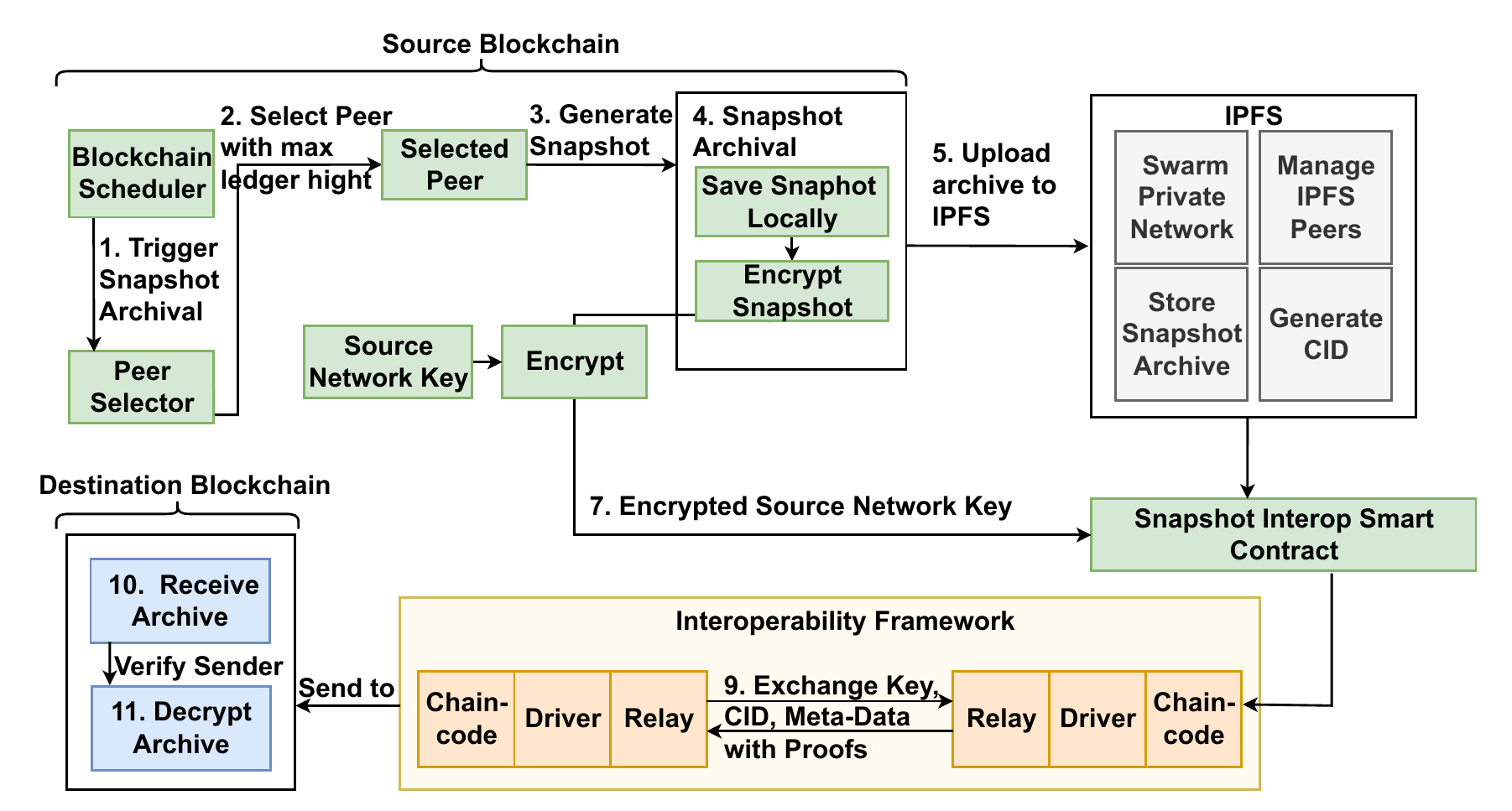}}
\caption{\ours Workflow for Cross-Network Snapshot Archive Transfer}
\label{fig:architecture}
\end{figure}

Firstly, \ours enforces receipts of every inter-blockchain transaction, capturing non-repudiable proof of all such events (Fig.~\ref{fig:crosschaintransaction}). Thereafter, it follows a workflow for saving, encrypting, archiving, and transferring snapshots across networks as depicted in Fig.~\ref{fig:architecture}.
The workflow starts with a need-based scheduler triggering a snapshot process when there is a significant number of new un-archived transactions and blocks in the ledger.
A peer is then selected among the peers having the maximum ledger height to generate a snapshot of the current blockchain state.
Since it is infeasible to fully verify one blockchain’s state without the entire historical lineage \cite{borkowski2018caught}, \ours explicitly includes
endorsements (in the form of signatures) of the blockchain participants on the cross-chain transactions, as well as the corresponding receipts for verifiability.
The snapshot is then encrypted with the source network’s encryption key and uploaded to a private IPFS network, producing a unique Content Identifier ($\mathtt{CID}$) for reference. 
The $\mathtt{CID}$, decryption key, and metadata are then transmitted to a different network through the interoperability framework\cite{cactis}, for dispute resolution and external auditing. 
The destination network verifies the $\mathtt{CID}$ and decrypts the archive 
with the provided key. Access to this snapshot data containing transaction receipts helps in dispute resolution.
In traditional permissioned blockchain frameworks, such as Hyperledger Fabric \cite{hlf}, 
snapshots are typically generated and stored locally \cite{hlfledgersnapshot}. However, the \ours architecture advances beyond this point to establish a mechanism for non-repudiable inter-chain transaction receipts, and a cross-network snapshot sharing process for enforcing accountability.
The subsequent sections detail the major functional components of \ours.

\subsection{Recording Non-Repudiable Transaction Receipt}
\label{subsec:receipts}
When a cross-chain transaction $T_{cc}$ is invoked from network $\mathcal{N}_1$ to network $\mathcal{N}_2$, different methods can be used for trustworthy transfer of data between blockchains, such as token exchanges, notary schemes, side-chains, inter-blockchain connectors, etc.~\cite{herlihy2018atomic, scheid2021notary, back2014enabling, dltinterop}. \ours uses Hyperledger Cacti \cite{cactis} for cross-chain transactions which utilizes endorsements by the participants of the sender network $\mathcal{N}_1$ in the form of digital signatures \cite{abebe2019enabling}. These endorsements allow the receiving network $\mathcal{N}_2$ to verify that $\mathcal{N}_1$ is indeed the sender, and participants of $\mathcal{N}_1$ have consensus on the data/asset it is sending across. However, neither Hyperledger Cacti nor Fabric has any provision to record the confirmation of whether $\mathcal{N}_2$ has successfully received and accepted the transaction. In order to ensure non-repudiability and prevent the destination network $\mathcal{N}_2$ from raising fraudulent disputes, a proof of 
the cross-network transactions has to be preserved. To ensure that $\mathcal{N}_2$ cannot deny the acceptance of the transaction,
we propose a receipt process for each cross-network transaction. 
\ours makes sure that the invoked and receipt transactions are clubbed under a transaction set $\mathcal{T}_{\text{Set}}$ as depicted in Fig. \ref{fig:crosschaintransaction}.
To maintain atomicity, when a cross-network transaction $T_{cc}$ is initiated from $\mathcal{N}_1$ to $\mathcal{N}_2$, the transaction set $\mathcal{T}_{\text{Set}}$ would be considered complete only when $\mathcal{N}_2$ sends a receipt transaction to  $\mathcal{N}_1$ in reply within a stipulated time limit. 
Thus, $\mathcal{N}_1$  can directly validate ${T_{rt}||\psi_{\mathcal{N}_2}(T_{rt})}$ for establishing the truth.
If the receipt transaction is not received beyond the time limit, the entire transaction set is marked as incomplete and excluded from future references. 

It is challenging to establish trust between $\mathcal{N}_1$ and $\mathcal{N}_2$, particularly in scenarios where mutual faith is absent. Data of any network $\mathcal{N}_i$ consisting of $|\mathcal{N}_i|$ peers is trusted by its peers if the data is majority endorsed ($|\psi_{\mathcal{N}_{i}}(T_{cc})| \geq \frac{2|\mathcal{N}_i|}{3}$) and passed through respective consensus rules. But a local network, upon receiving the external data from a foreign network, only knows the data must have been endorsed by the majority of peers of that network, but the networks may not trust each other inherently. To ascertain the local network's confidence in the received information, we advocate for the services of an independent auditor.
This methodology aims to provide a mechanism for verifying cross-chain transactions and resolving disputes effectively.
The cross-chain snapshot archives are sent to an auditor on a regular basis by both networks whenever a snapshot is generated in either of the participating networks. For example, when a network $\mathcal{N}_1$ generates a snapshot of its current state containing transactions, metadata, hashes, etc., the snapshot is encrypted and then sent to the auditor. The same approach is followed by $\mathcal{N}_2$ for sending snapshots to the auditor. This ensures that the auditor always has the latest state of both networks. The auditor is assumed to be a reputed one and can be trusted by both of them. If either of the networks does not believe in the claim by the other network, it can invoke the auditor service to verify whether the transaction really happened before. The auditor service will search the transaction in pre-existing cross-snapshot archives. If a matching entry is found, the auditor will send its consent, and the networks will unanimously agree to that, settle their claims, and maintain the records in respective ledgers. If no such valid transaction is found by the auditor, the claim by the demanding network is refuted without any further disputes. 
\ours introduces cross-chain sharing of snapshots in order to facilitate such external audit process as described in the following sub-sections.  

\subsection{Need-based Snapshot Scheduling}
\label{subsec:needbasedschedule}
\begin{figure}[!ht]
\centering
\centering
{\includegraphics[scale=0.35]{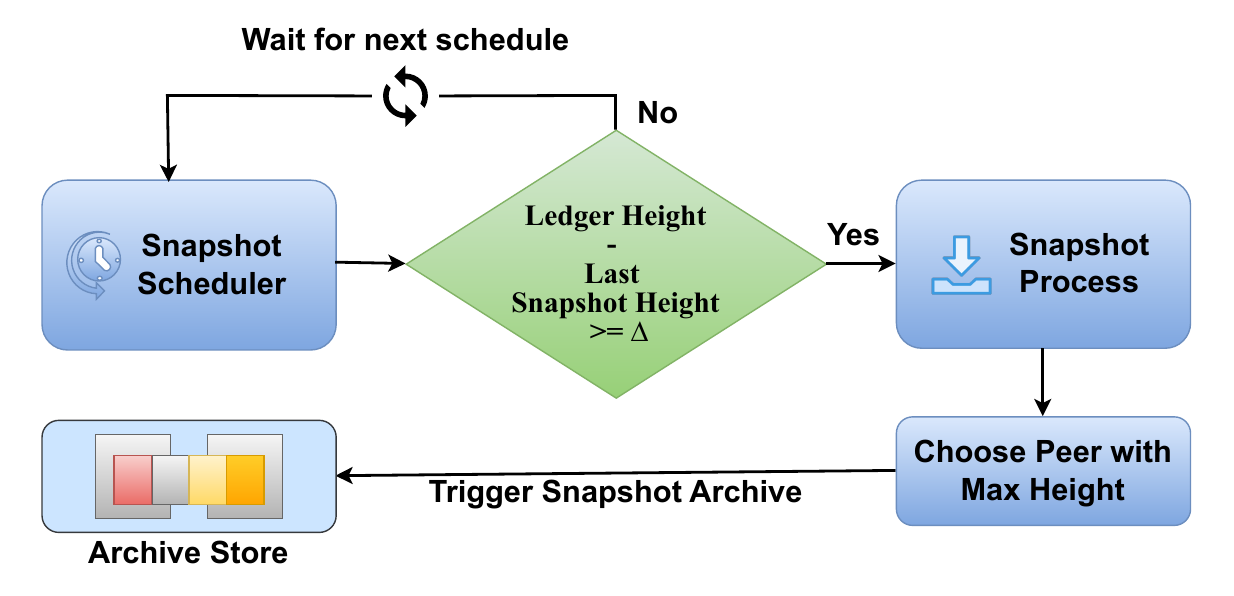}}
\caption{\notets{Need based Snapshot Archival}}
\label{fig:needbasedsnapshotarchive}
\end{figure}

Permissioned blockchains often offer scheduled snapshot features triggered when the ledger reaches a predefined height \cite{hlc}. However, this approach lacks flexibility as a newly generated snapshot overwrites previous snapshot data, preventing to trace back to some earlier system state in the event of sudden crashes or peer malfunctions. 
To address this challenge, we introduce an algorithm (Algorithm \ref{algo:1}) that automatically triggers the snapshot archival process upon detection of significant ledger changes from the last snapshot, based on a threshold parameter $\Delta$.
We set this value as $ G \times T $, where $G$ is the average number of blocks added per hour to the ledger and  $T$ is the time (in hours), which is a tunable parameter. For example, if $T$ is set to 24, that implies 24 hours worth of new blocks after the last snapshot will trigger a new snapshot. 
The algorithm's core logic is encapsulated in the function \texttt{ProcessSnapshot} (refer to Algorithm \ref{algo:1}). This function evaluates whether the difference between the current ledger height and the last recorded snapshot height exceeds the threshold \( \Delta \). \notets{If this condition is satisfied, it triggers a snapshot archival process
as depicted in Fig. \ref{fig:needbasedsnapshotarchive}}.
The function is invoked periodically every \( h \) hours. This periodic invocation ensures the significant change of ledger state is regularly assessed, optimizing the snapshot archival process based on ledger activity.
\RestyleAlgo{ruled}
\SetKwComment{Comment}{/* }{ */}
\SetKwInput{KwInput}{Initialization}
\SetKwFunction{ProcessSnapshot}{ProcessSnapshot}

\begin{algorithm}[hbt!]
\footnotesize
\caption{Need-based Snapshot Scheduling}
\label{algo:1}

\KwInput{
Snapshot Height \( |S_{\mathcal{N}_r}| \gets -1 \)\;
Threshold \( \Delta \gets G \times T \)\;
Ledger Height \( |\mathcal{L}_{\mathcal{N}_r}| \)\;
}

\vspace{0.2cm}

\SetKwProg{Fn}{Function}{:}{}
\Fn{\ProcessSnapshot{\(|S_{\mathcal{N}_r}|, |\mathcal{L}_{\mathcal{N}_r}|, \Delta\)}}{
    \If{\(|\mathcal{L}_{\mathcal{N}_r}| - |S_{\mathcal{N}_r}| \geq \Delta\)}{
        \textbf{Invoke snapshot archival}\;
        \( |S_{\mathcal{N}_r}| \gets |\mathcal{L}_{\mathcal{N}_r}| \)\;
    }
    
}

\textbf{End} Function

\vspace{0.2cm}

\textbf{Every $h$ hours periodically invoke}:\\
\Indp \textbf{Call} \ProcessSnapshot{\(|S_{\mathcal{N}_r}|, |\mathcal{L}_{\mathcal{N}_r}|, \Delta\)}\
\end{algorithm}

\subsection{Selection of Snapshot Generating Peer}
\label{subsec:choosepeer}
Selecting an appropriate peer say $p_{i}$, to capture a ledger snapshot is crucial in a permissioned blockchain network. 
Various peer selection strategies can be opted for, including selecting a peer randomly, choosing the most recently created peer, or selecting the peer with the maximum historical references. However, random selection may yield unsatisfactory results by picking peers with limited transaction history, risking omitting crucial transactional details during new peer bootstrapping. Similarly, selecting the most recent peers or newly created ones may pose risks due to their readiness status, as they may only be fully operational once routine peer validations are completed, or they may not possess complete chronological references.
In \ours we choose the peer with maximum ledger height for snapshot generation.
If two peers have the same height, one of the peers will be chosen randomly for snapshot generation.
We determine the maximum ledger height as 
{$|\mathcal{L}_{\mathcal{N}_1}|= \mathsf{Max(|\mathcal{L}_{\mathcal{N}_1}^{p_{1}}|),|\mathcal{L}_{\mathcal{N}_1}^{p_{2}}|, \dots,|\mathcal{L}_{\mathcal{N}_1}^{p_{i}}|,\dots,|\mathcal{L}_{\mathcal{N}_1}^{p_{|\mathcal{N}_{1}|}}|)}$, where $|\mathcal{L}_{\mathcal{N}_1}^{p_{i}}|$ represents the ledger height of $p_i$, and $ |\mathcal{N}_{1}|$ represents the total number of peers in $\mathcal{N}_{1}$.
} Unlike traditional approaches that rely on static query configurations established during bootstrap, \ours uses the \textit{Service Discovery Protocol} (SDP) \cite{hlfsdp} that dynamically retrieves the network topology and peer details, bypassing dependencies on direct peer communications. Thus, the risks of manipulation by malicious peers are mitigated.
We apply this protocol to discover the peer with the maximum ledger height in the permissioned blockchain.
\subsection{Archive Encryption}
\label{subsec:archiveencryption}
In this subsection, we describe the encryption process used to preserve the confidentiality of snapshot archives. Since these snapshots can grow in size and contain sensitive state and endorsement data, we adopt a symmetric encryption scheme optimized for performance. The goal is to ensure security and low storage occupancy without significantly impacting throughput.
Notably, snapshot archives contain sensitive information, including transaction histories, organizational schema, and membership details. To protect their confidentiality, encryption is essential, especially when saving data in the decentralized storage or transferring them to external auditors. By encrypting the snapshots, the risk of exposure or tampering in decentralized systems is mitigated.
In permission blockchains, such as Hyperledger Fabric, the latest snapshots are kept in a peer directory in a signed form, and only administrators can access them, but there is no provision for archiving the snapshots. To overcome the challenge, we propose an encryption mechanism to securely store the data. 
\ours uses GPG (GNU Privacy Guard)
\footnote{\url{https://www.gnupg.org} (Accessed: \today)}
as the encryption tool. It is an open-source tool popularly used for file encryption. Cryptographic functions and keys are utilized by GPG to facilitate data encryption.
GPG employs 128-bit AES cipher \cite{daemen1999aes} symmetric key encryption. We use a random passphrase as input to GPG. GPG uses this passphrase to derive a symmetric key using a key derivation function (KDF). This KDF incorporates a random salt to the function \cite{pgprfc}. GPG hashes the salt and passphrase multiple times using SHA 2 to ensure that even if two people use the same passphrase, they will end up with different keys. The SHA 1, though still used, is getting outdated, and SHA 3 is slower and does not have native support for GPG, so we select the stable SHA 2 (512 bits)\footnote{\url{https://csrc.nist.gov/pubs/fips/180-4/upd1/final} (Accessed: \today)}.
GPG then uses the AES algorithm to encrypt the data using the derived key. 
The produced encrypted snapshot archive is saved and shared on private IPFS \cite{ipfsdc}.
As the size of snapshot archives can grow exponentially with time, the symmetric key is preferred over the asymmetric key cryptography \cite{diffie1976new}. 
It is technically possible to use asymmetric public key encryption as well, however, it is computationally more intensive to handle general purpose encryptions \cite{alvarez2016algorithms}. So we used a random symmetric key for encrypting snapshot archives.
The key is also saved within the participating organization's wallet as a backup so that in case of ledger failure, the data can be recovered from encrypted storage. 
When there is a need to share the snapshot archive with an external network, this key is exchanged confidentially through a cross-network transaction. This enables the destination network to decrypt and verify the archive content.
\subsection{Saving Encrypted Archives in Decentralized Storage}
\label{subsec:encryptedipfs}
Although any secure bulletin board or distributed ledger could be used to store snapshot data, we choose the InterPlanetary File System (IPFS)~\cite{ipfsdc} for its scalability and decentralization benefits. IPFS is a widely used distributed storage system that employs peer-to-peer protocols for address mapping, routing, and exchanging content-addressed data. As blockchain ledgers grow in size, storing complete snapshot archives directly on a public blockchain such as Ethereum becomes prohibitively expensive due to gas fees and storage overhead. In contrast, off-chain storage with IPFS avoids these costs, since the blockchain only needs to store a reference to the snapshot, which is its Content Identifier ($\mathtt{CID}$). Once the snapshot archive is uploaded to IPFS, a unique $\mathtt{CID}$ is generated and can be recorded on-chain with minimal overhead.
In \ours architecture, the snapshot archive is stored in a compressed format to further conserve storage space, then loaded into IPFS using a permissioned blockchain peer from the source network. 
The archive content cannot be retrieved without knowledge of the unique \( \mathtt{CID} \). So,
$
\mathtt{CID} = H(C_{arc}) = H(E_k(
{S_{arc}})) = H(E_k([S^{(1)}  \cup S^{(2)},\ldots, S^{(m)}]))$,
where $H(.)$ represents the cryptographic hash function, $S^{(m)}$ is the snapshot at ${m}^{th}$ timestamp,  $S_{arc}$ is the snapshot archive,
$C_{arc}$ represents encrypted snapshot archive, and $E_{k}$ is encryption $E$ with the encryption key $k$.
Therefore, the $\mathtt{CID}$ is the hash reference of the collection of the encrypted snapshots. A malicious user cannot modify the original archive content without altering the $\mathtt{CID}$. Furthermore, retrieving the snapshot archive requires the encryption key, which is not even uploaded on IPFS. Thus, both avenues ensure the safety and confidentiality of the snapshot contents.
Although the encrypted snapshot archives uploaded to IPFS are inherently secure, further enhancement of end-to-end privacy in snapshot delivery is achieved by utilizing a private IPFS network in the design,
thereby restricting access to only those peers that were chosen to exchange the encrypted snapshot archive files.
\subsection{Data Interoperability}
\label{subsec:hlfcrossnet}
\begin{figure}[!t]
    \centering
    \includegraphics[width=0.86\linewidth]{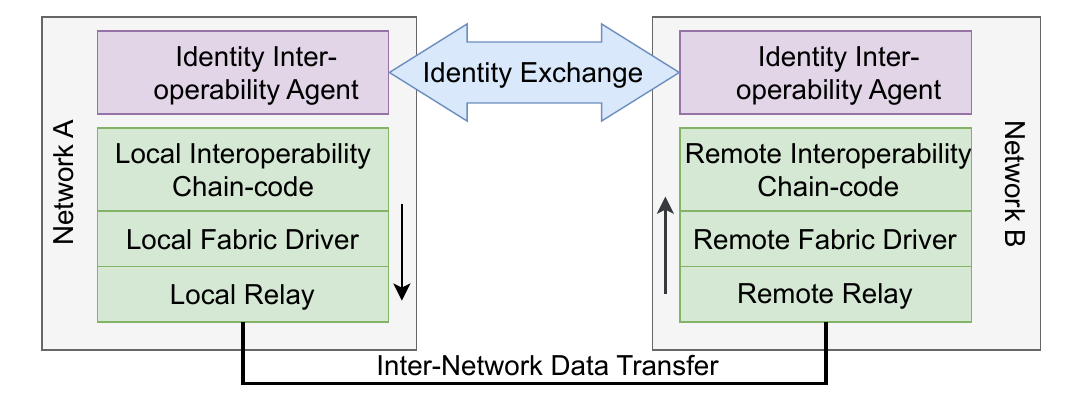}
    \caption{Layered Inter-Network Communication Diagram}
    \label{fig:layeredinterconnection}
\end{figure}

We next discuss how \ours, combined with IPFS, relay drivers, smart contracts, and identity exchange protocols, works to ensure end-to-end verifiability and trust between fabric networks.
For external audit or fault recovery purposes, the archives need to be transferred to the foreign networks.
After the archive data is saved on the private IPFS network, the $\mathtt{CID}$ of that encrypted archive, and decryption key are together transferred from the source network to the destination network. The $\mathtt{CID}$ is needed to identify and extract the archived file contents from IPFS at the destination network. The decryption key is used to decrypt the encrypted archive as depicted in Fig.~\ref{fig:architecture}. 
The \ours architecture integrates a permissioned blockchain interoperability framework that uses \textit{relays} and proof mechanisms with attestations, to establish cross-chain trust, as introduced by Abebe et al \cite{abebe2019enabling}. This protocol is further fine tuned in the Hyperledger Cacti project \cite{cactis}. As depicted in Fig.~\ref{fig:layeredinterconnection},  
\textit{relays} act as a gateway for communication between the disparate networks. The \textit{relay} is a ledger independent module that handles data transfer through a blockchain agnostic protocol. These \textit{relays} are hooked to a \textit{driver} module that facilitates coordination between the blockchain ledger and the \textit{relay}. For handling interoperation functionality, each ledger has its own smart contract referred to as \textit{interoperability smart contract} (\textit{interoperability chaincode} in Fabric). When a data is to be transferred from the source blockchain to the destination blockchain, the \textit{interoperability smart contract} first initiates the cross-network transaction. It prepares the data and attaches the proof of consensus on the data in the form of attestations using digital signatures of the consensus participants. The \textit{driver} converts this data to a network agnostic format and passes it to the \textit{relay}. Through the \textit{relay} and then the \textit{driver} of the destination network, this data is received by the  \textit{interoperability smart contract} of the destination. The destination network validates the attestation to ensure veracity of the data. Notably, the validation of attestations requires the identity information of the foreign network. For this, an identity interoperability protocol is used \cite{ghosh2021decentralized} through which identity interoperation network agents (IIN Agents) exchange the identity of different blockchain network participants' identities and certificates.
In order to preserve confidentiality of the $\mathtt{CID}$ and the decryption key, \ours encrypts the contents using the public key of the destination network. Furthermore \ours enforces a receipt transaction against this inter-network transfer also. 

\section{Security Analysis}
\label{sec:analysis}
This section provides a detailed security analysis of the proposed architecture, focusing on a line of defense against various types of attacks while ensuring the essential properties of distributed interoperable systems.

\subsection{Types of Attacks Defended Against}

\noindent \textbf{Collusion Attack}: Participants from one network ($\mathcal{N}_1$) may collude together to send fraudulent transactions to another network ($\mathcal{N}_2$), and later deny them (refer to Sec. \ref{subsec:trustmodel}). \ours keeps all such cross-chain transactions along with data and proofs in the form of attestations of $\mathcal{N}_1$'s participants in the snapshot. The snapshots are then transferred to an external auditor network, which can clearly find evidence of $\mathcal{N}_1$'s transactions.

\noindent \textbf{Fraudulent Demands}: 
A malicious network $\mathcal{N}_1$, in spite of receiving data/asset successfully from $\mathcal{N}_2$ may refuse to acknowledge the same. The rogue network $\mathcal{N}_1$  may falsely claim that it never received the data/asset. To counter this, \ours ensures all cross-chain transactions are verified via \textit{Transaction Receipts} (refer to Sec. \ref{subsec:receipts}). A cross-chain transfer from $\mathcal{N}_2$ to $\mathcal{N}_1$ is completed only when a receipt transaction is signed and sent back from $\mathcal{N}_1$ to $\mathcal{N}_2$. As a result, $\mathcal{N}_2$ has undeniable proof that $\mathcal{N}_1$ had received the concerned data/asset.

\noindent \textbf{Ledger Fault}: In cases of network failures that result in ledger data loss, malicious actors could exploit incomplete data to create fraudulent claims. To keep snapshots up to date, \ours selects peers for archival based on the currency of their ledger data. Peers that are intentionally or unintentionally trailing behind the latest ledger height are not chosen for archival.
The architecture ensures that all snapshot archives are stored securely in a decentralized manner using \textit{IPFS}.

\subsection{Safety and Liveness  Guarantees}
The architecture ensures several key security guarantees through a combination of encryption, private IPFS communication, and receipt transactions. Here we discuss the safety and liveness guarantees supported by \ours.

\subsubsection{\textbf{Safety}}
\ours ensures safety by enforcing atomicity between a cross-chain transaction and its corresponding receipt. The transaction set $\mathcal{T}_{\text{Set}}$ is considered complete only when both $T_{cc}$ and $T_{rt}$ are finalized, guaranteeing that the transaction and the receipt have been endorsed by a majority of peers in both networks. Even if one network later attempts to dispute the cross-chain exchange, the stored endorsements and receipt provide verifiable proof of the original consensus. Therefore, under the fault model presented in Section~\ref{subsec:faultmodel}, and assuming the underlying ledgers are not Byzantine faulty, \ours preserves safety and remains resilient against fraudulent claims and ledger data loss. We next elaborate on the Integrity, Confidentiality, Authenticity, and Non-repudiation aspects of the safety requirements.

\begin{itemize}

    \item \textbf{Integrity:} Integrity is maintained by ensuring that the destination network receives the cross-chain transaction content exactly as sent. This is achieved through a unique Content Identifier (\texttt{CID}), generated by hashing the encrypted snapshot archive stored in IPFS, which guarantees that any data retrieved from IPFS remains unaltered. 
    
    \item \textbf{Confidentiality:} Confidentiality is ensured by encrypting all snapshot archives with AES-128 before storing them in the IPFS network. Access to the private IPFS network is restricted to authorized participants possessing the randomly generated swarm key. For snapshot sharing, the Content ID and decryption key are further encrypted with the receiver’s public key and transmitted through the interoperability framework, preserving confidentiality throughout the communication channel.
    
    \par \textit{Key Management:} Each snapshot is encrypted with a unique, randomly generated symmetric AES-128 key, created by the selected snapshot-generating peer (Section~\ref{subsec:choosepeer}). After the snapshot is shared via IPFS, the \texttt{CID} and key are committed to the ledger, allowing all participants to verify the key by accessing the snapshot. Once verified, each participant stores a backup of the key in an off-chain wallet. Although the key is generated by a single peer, it is shared with and agreed upon by all blockchain participants. Each encryption key is used for only one snapshot and is never reused.
    
    \item \textbf{Authenticity:} Endorsements in the form of signatures for each cross-chain transaction ensure the sender’s authenticity. This is reinforced by requiring a signed response from the destination network, confirming that the transaction has been received and acknowledged by the intended recipient.

    \item \textbf{Non-repudiation:} Once a cross-chain transaction $\mathcal{T}_{cc}$ and its corresponding receipt $\mathcal{T}_{rt}$ are endorsed and archived, neither party can repudiate them, as shown in the following proof.

        \fbox{%
        \begin{minipage}{0.96\linewidth}
        \small 
        
        \begin{IEEEproof}
        Assume that $\mathcal{N}_2$ (the destination network) attempts to repudiate the receipt of a cross-chain transaction $\mathcal{T}_{cc}$ initiated by $\mathcal{N}_1$ (the source network).
        
        \textit{Case 1:} A valid receipt $\mathcal{T}_{rt}$ exists, endorsed by a majority of peers in $\mathcal{N}_2$ and included in the snapshot archive. As this receipt is signed and cryptographically verifiable, repudiation by $\mathcal{N}_2$ contradicts its own prior endorsements and is therefore 
        not substantiated.
        
        \textit{Case 2:} No receipt $\mathcal{T}_{rt}$ was issued by $\mathcal{N}_2$. In this case, the transaction set $\mathcal{T}_{\text{Set}} = \{\mathcal{T}_{cc}, \mathcal{T}_{rt}\}$ is marked incomplete by \ours's protocol (see Section~\ref{subsec:receipts}). Thus, $\mathcal{N}_1$ cannot claim the transaction was acknowledged, and no repudiation arises.
        
        In both cases, repudiation is either provably false or structurally impossible. Therefore, \ours enforces non-repudiation by design.
        \end{IEEEproof}
        \end{minipage}}
\end{itemize}

\smallskip 
\subsubsection{\textbf{Liveness}}
Liveness denotes the system’s ability to make continual progress despite the presence of faults or failures. In \ours, this property is preserved by ensuring that the snapshot archival mechanism remains functional even when certain snapshot-generating peers are unavailable. This resilience is achieved through a distributed peer-selection process and a need-based snapshot scheduling policy, which triggers archival when  
\[
|\mathcal{L}_{\mathcal{N}_r}| - |S_{\mathcal{N}_r}| \geq \Delta
\]
(see Section~\ref{subsec:needbasedschedule}). Two primary operational cases are considered:  
\begin{enumerate}
    \item[I)] \textbf{Snapshot exchange via Hyperledger Cacti relays:}  
    Provided that the relays and driver modules are active, transactions and receipts will eventually propagate across networks, enabling cross-chain synchronization.
    \item[II)] \textbf{Persistent storage in IPFS:}  
    Snapshots are permanently archived in IPFS. Consequently, even if one or more networks crash, the stored snapshots remain accessible. Upon network recovery, these snapshots facilitate the resumption of operations and ongoing progress.
\end{enumerate}

In both cases, as long as the underlying blockchain and communication layers remain operational, \ours will eventually make progress. Beyond the core protocol design, liveness is maintained at the following operational levels:  
\begin{itemize}
    \item \textbf{End-Users:}  
    Atomicity is enforced between each cross-chain transaction $\mathcal{T}_{cc}$ and its corresponding receipt $\mathcal{T}_{rt}$, treating them as a single transaction set. A predefined timeout ensures that incomplete sets are discarded, guaranteeing that user-initiated transactions are eventually either committed or rejected.
    \item \textbf{Ledgers:}  
    \ours does not interfere with the intrinsic liveness of participating blockchains. Each network continues to accept, validate, and commit transactions independently. Receipt enforcement persists as long as the underlying ledgers remain live; incomplete transaction sets are timed out without hindering others.
    \item \textbf{Auditors:}  
    Auditors retrieve snapshot archives from IPFS using the shared decryption key, swarm key, and Content ID. These items are exchanged via the interoperability framework (e.g., Hyperledger Cacti), whose liveness depends on the blockchain and communication modules. Auditing is performed off-chain, ensuring that it does not impact blockchain or inter-blockchain liveness.
\end{itemize}

Collectively, these mechanisms enable \ours to operate reliably in the presence of faults such as missing receipts or the failure of snapshot-generating peers, thereby upholding its liveness guarantees.

\section{Implementation Details}
\label{sec:implemenation}
We have developed a proof-of-concept implementation of \ours, designed to capture cross-blockchain transactions together with their non-repudiable receipts. The implementation is open-sourced and available to the community\footnote{ \url{https://github.com/mailtisen/InterSnap} (Last accessed: \today)}. We create two consortium HLF networks (A and B as shown in Fig. \ref{fig:network}). The subsequent sub-sections describe the implementation details and the configuration of the test blockchain network.
\subsection{Archive Snapshot }
We use the Hyperledger Cacti framework to implement receipts for cross-blockchain transactions. 
The interoperability chain code ensures that a cross-chain data/asset transfer transaction is completed only when the destination network sends a receipt with requisite signatures. Transaction data and their non-repudiable receipts are captured in the snapshot archive. 
We run the \texttt{getdiscoveryMaxPeerHeight} program to find the peer with maximum height so that the peer is chosen for snapshot generation. We periodically use  Algorithm ~\ref{algo:1} to trigger a snapshot archival process when there has been a significant change in the ledger data.
\subsection{Secure Transfer of Archives through Private IPFS}
\begin{figure}[!ht]
\centering
\includegraphics[scale=0.16]
{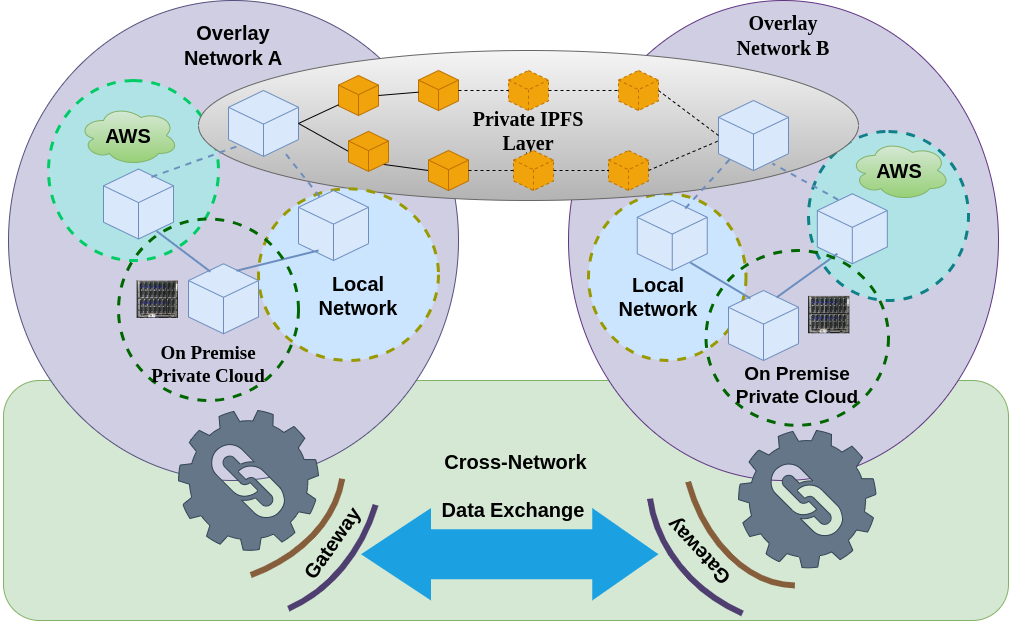}
\caption{Overlapping Network Diagram Between Hyperledger Fabric and Private IPFS}
\label{fig:network}
\end{figure}

We compress the snapshot archives using the tar 
\footnote{\url{https://man7.org/linux/man-pages/man1/tar.1.htm} (Accessed: Aug 6, 2025)} tool, and encrypt them with GPG as described in \ref{subsec:archiveencryption}.
We use Go kubo (version 0.17.0) distribution for IPFS for uploading archives to IPFS. For each encrypted snapshot archive upload, IPFS returns a unique $\mathtt{CID}$.
Next, we launch the IPFS daemon to initialize the bootstrap node. We generate and set the swarm key in the specific machines and launch them as peer nodes to establish the private IPFS network. We run the \texttt{uploadArchiveToIPFS} program to upload the encrypted and compressed snapshot archives in private IPFS.
as depicted in Fig.~\ref{fig:network}. 
\subsection{Interoperable Snapshot Archive Transmission}
We deploy custom chain codes on both networks which invoke outward and inward services of the Hyperledger Cacti framework~\cite{cactis} 
through the relay. We also implement a time-based locking mechanism for every cross-network asset exchange or transaction. This ensures that the invoking $T_{cc}$ and receipt transaction $T_{rt}$ are treated as a single atomic transaction set $\mathcal{T}_{\text{Set}}$. If both $T_{cc}$ and $T_{rt}$ are successful, they are recorded in $\mathcal{L}_{\mathcal{N}_1}$ and $\mathcal{L}_{\mathcal{N}_2}$, ensuring that snapshots contain both entries. 
The interoperable relay as an intermediary service 
invokes the gRPC\footnote{\url{https://grpc.io/} (Accessed: \today)}, based lightweight message exchange protocol to exchange payload between the two HLF networks.
On top of the Fabric Relays, we set up our Fabric Drivers. 
We set up a basic Interoperable Identity Network (IIN) layer to synchronize certificates with the other network using the IIN agents. 
The data exchange happens with multiple stages as mentioned in Fig.~\ref{fig:layeredinterconnection}, such as local peer-to-local network (by local interoperability program and local driver service), then the local network to the remote network (by relay service), and finally, the remote network to remote destination peer (by remote driver and remote interop chain code). The membership details are securely exchanged by the inter-network agents as well for signed and authenticated responses.
On the other hand, The peer node of Network B downloads the snapshot archive using the received CID. The transferred key is used to decrypt the snapshot archive at Network B. Along with that, the interoperability setup enables a secure exchange of metadata across networks, which is required for maintaining confidentiality and integrity during snapshot sharing.

\subsection{Integrating Other Blockchain Platforms}
While \ours is currently implemented using Hyperledger Fabric and Hyperledger Cacti, its architectural design is inherently adaptable to other interoperability frameworks such as Cosmos IBC~\cite{kwon2019cosmos}, Chainlink CCIP~\cite{breidenbach2021chainlink}, and Polkadot XCMP~\cite{wood2016polkadot}. For instance, in Cosmos, the SDK provides a native snapshot mechanism called \emph{state sync}, primarily used for efficient node synchronization. This mechanism captures the complete application state at a specific block height and delivers it in compressed chunks to new nodes joining the network. In the Cosmos IBC protocol, cross-chain communication is secured through proofs over Tendermint~\cite{buchman2016tendermint} consensus. In this context, the endorsement logic in \ours could be replaced with IBC proofs of commitment (i.e., Merkle roots of state changes). Similarly, in Chainlink CCIP, trusted oracle signatures could serve as a substitute for Fabric’s endorsements to verify transaction validity. In the case of Polkadot XCMP, snapshot generation could be integrated into the relay chain, leveraging its message-exchange framework to enforce snapshot exchange. We plan to investigate the adaptability of \ours to these and other interoperability protocols in future work.

\section{Experimental Results}
\label{sec:results}
To evaluate our proposed \ours architecture, we have experimented extensively with respect to various parameters including varying ledger height and number of archived transactions in a snapshot. 
\subsection{Test bed Set up and Data Preparation}
The tests are carried out using four on-premises physical machines with Intel® Core™ i5-4570 CPU, $8$ GB RAM, and $64$-bit Ubuntu 18.04.6 LTS operating system. Along with these, we also include three AWS EC2 cloud instances, and one virtual machine from the on-premises private cloud infrastructure of similar configuration. 
We have bridged the communication path between these networks to build the overlay network using the docker swarm \footnote{\url{https://docs.docker.com/engine/swarm/} (Accessed: \today)}.
The Hyperledger Fabric (release 2.4) networks were run on the said Ubuntu environment with block size set to $512$ KB. The private IPFS network nodes were also set up. 
To prepare the snapshot data, we take snapshots at various stages: during the initial network launch, after adding a new organization, and after new transaction commits. 
After each significant data increment, our snapshot archive program checks, generates, and stores snapshots for further processing.
We varied the transaction payload and content within a block by introducing delays between submissions, submitting multiple transactions in quick succession, updating multiple keys simultaneously, and changing the data size. This approach creates a diverse snapshot mix for comprehensive testing.

\begin{figure}[!ht]
\begin{minipage}{0.48\linewidth}
\centering
\begin{tikzpicture}[scale=0.46]
    \begin{axis}[
        xlabel={\textbf{No. of Archived Transactions (X times $1000$)}},
        ylabel={\textbf{Avg Encryption Time (in seconds)}},
        ymin=0.4, ymax=0.7,
        xmin=0, xmax=12,
        xtick={0, 2, 4, 6, 8, 10, 12, 14, 16},
        ytick={0,0.4, 0.45, 0.5, 0.55, 0.6, 0.65},
        legend pos=north west,
        ymajorgrids=true,
        grid style=dashed,  
    ]
    
    \addplot[color=green, mark= diamond]
        coordinates {(1, 0.456)(2, 0.458)(5, 0.457) (8, 0.458) (10, 0.468)(12, 0.461)};
        \addlegendentry{Ledger Height $1K$}

   \addplot[color=violet, mark= square]
        coordinates {(1, 0.489)(2, 0.493)(5, 0.494) (8, 0.495)(10, 0.495)(12, 0.494)};
        \addlegendentry{Ledger Height $5K$}
        
  \addplot[color=orange, mark= triangle]
        coordinates {(1, 0.535)(2, 0.534)(5, 0.537) (8, 0.540)(10, 0.541)(12, 0.544)};
        \addlegendentry{Ledger Height $10K$}
    \end{axis}
    \end{tikzpicture}
    \caption{Snapshot Archive Encryption Time}
    \label{fig:snaparc_encr}
\end{minipage}
\hfill 
\begin{minipage}{0.48\linewidth}
\centering
\begin{tikzpicture}[scale=0.46]
    \begin{axis}[
        xlabel={\textbf{No. of Archived Transactions (X times $1000$)}},
        ylabel={\textbf{Save to IPFS Time (in seconds)}},
        ymin=0, ymax=0.7,
        xmin=0, xmax=12,
        xtick={0, 2, 4, 6, 8, 10, 12, 14, 16},
        ytick={0,0.1, 0.2, 0.3, 0.4, 0.5, 0.6},
        legend pos=north west,
        ymajorgrids=true,
        grid style=dashed,  
    ]
    
    \addplot[color=green, mark= diamond]
        coordinates {(1, 0.325)(2, 0.35)(5, 0.321)(8, 0.324) (10, 0.329) (12, 0.332)(15, 0.328)};
        \addlegendentry{Ledger Height $1K$}

   \addplot[color=blue, mark= square]
        coordinates {(1, 0.35)(2, 0.374)(5, 0.356) (8, 0.371) (10, 0.366)(12, 0.339)(15, 0.351)}; 
        \addlegendentry{Ledger Height $5K$}
        
  \addplot[color=orange, mark= triangle]
        coordinates {(1, 0.458)(2, 0.448)(5, 0.444) (8, 0.464) (10, 0.434) (12, 0.401)(15, 0.445)};
        \addlegendentry{Ledger Height $10K$}
    \end{axis}
    \end{tikzpicture}
    \caption{Saving Encrypted Archive to IPFS}
    \label{fig:snaparc_up_ipfs}
\end{minipage}
\end{figure}
\subsection{Snapshot Latency}
We test and compare the encryption time for archived transactions below. Fig. \ref{fig:snaparc_encr} represents the average time taken for encrypting the archives containing varying numbers of transactions. 
The figure shows that the encryption time trend of the archives does not deviate much with the increment of the number of archived transactions. Even for snapshots with $12,000$ transactions, the encryption takes less than $0.6$ seconds.
To study the overall latency of saving archives to IPFS, we vary the total number of transactions in a snapshot. 
We observe that the average time taken to save the snapshot archives to IPFS is steady with the varying number of archived transaction content as depicted in Fig. \ref{fig:snaparc_up_ipfs}. \ours completes encryption and saving to IPFS, together under a second for a snapshot archive containing $12,000$ transactions.  

\begin{figure}[!ht]
\begin{minipage}{0.48\linewidth}
\centering
\begin{tikzpicture}[scale=0.49]  
\begin{axis}  
[  
    ybar,  
    enlargelimits=0.15,  
    ylabel={\ \textbf{Average Snapshots per minute}}, 
    xlabel={\ \textbf{Ledger Height}},  
    symbolic x coords={1000, 5000, 10000, 15000, 20000}, 
    {ymin=50,ymax=115},
    xtick=data,  
     ,  
    nodes near coords align={vertical}, 
    legend pos=north west,
     ymajorgrids=true,
     grid style=dashed,
    ]  
\addplot [draw=blue, fill=cyan]coordinates {(1000,110) (5000,115) (10000,105) (15000,103) (20000,106)};  

\end{axis}  
\end{tikzpicture}
\caption{Snapshot Archive Throughput}
\label{fig:arcthroughput}
\end{minipage}
\hfill 
\begin{minipage}{0.48\linewidth}
\centering
\begin{tikzpicture}[scale=0.49]
    \begin{axis}[
        xlabel={\textbf{No. of Archived Transactions (K)}},
        ylabel={\textbf{Snapshot Generation Time (in seconds)}},
        ymin=0, ymax=1,
        xmin=0, xmax=10,
        xtick={0, 2, 4, 6, 8, 10, 12, 14, 16},
        ytick={0,0.1, 0.2, 0.3, 0.4, 0.5, 0.6, 0.7, 0.8},
        legend pos=north west,
        ymajorgrids=true,
        grid style=dashed,  
    ]

  \addplot[color=blue, mark= square]
    coordinates {(1, 0.504)(2, 0.532)(3, 0.526)(4 0.502)(5, 0.506) (6, 0.522) (7, 0.499)(8, 0.507) (9, 0.509) (10, 0.505)}; 
    \addlegendentry{Ledger Height $1K$}
        
    \addplot[color=green, mark= diamond]
        coordinates {(1, 0.547)(2, 0.55)(3, 0.541)(4 0.543)(5, 0.576) (6, 0.52) (7, 0.522)(8, 0.491) (9, 0.563) (10, 0.574)};
        \addlegendentry{Ledger Height $5K$}

  \addplot[color=brown, mark= triangle]
        coordinates {(1, 0.598)(2, 0.538)(3, 0.545)(4 0.577)(5, 0.565) (6, 0.543) (7, 0.575)(8, 0.56) (9, 0.582) (10, 0.591)};
        \addlegendentry{Ledger Height $10K$}

   \addplot[color=red, mark= star]
        coordinates {(1, 0.436)(2, 0.444)(3, 0.435)(4 0.445)(5, 0.434) (6, 0.421) (7, 0.45)(8, 0.44) (9, 0.436) (10, 0.437)};
        \addlegendentry{Fabric As-Is Snapshot}      
    \end{axis}
    \end{tikzpicture}
    \caption{Avg. Snapshot Generation Time}
    \label{fig:snapshotgeneration}
\end{minipage}
\end{figure}

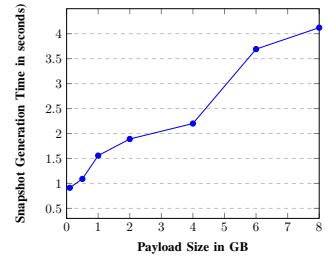
\begin{figure}[h!]
    \centering
    \begin{tikzpicture}[scale=0.49]
        \begin{axis}[
            xlabel={\textbf{Payload Size in GB}},
            ylabel={\textbf{Snapshot Generation Time in seconds)}},
            ymajorgrids=true,
            xmajorgrids=false,
            grid style={dashed},
            mark size=2pt,
            legend style={at={(0.5,-0.15)}, anchor=north},
            xtick={0,1,2,3,4,5,6,7,8},
            ytick={0.5,1.0,1.5,2.0,2.5,3.0,3.5,4.0},
            xmin=0, xmax=8,
            ymin=0.3, ymax=4.5,
        ]
        \addplot[
            color=blue,
            mark=*,
            thick
        ] coordinates {
            (0.100, 0.913)
            (0.500, 1.091)
            (1, 1.558)
            (2, 1.89)
            (4, 2.20)
            (6, 3.69)
            (8, 4.12)
            
        };
        \end{axis}
    \end{tikzpicture}
    
    \caption{Snapshot Time vs Payload Size}
    
    \label{fig:snapshotpayloadgeneration}
\end{figure}

\color{black}


\subsection{Snapshot Throughput}
Next, we measure the throughput of the 
snapshot generation
in terms of snapshots per minute. 
When generating a snapshot archive for a blockchain ledger, the number of transactions contained within the archive is directly influenced by the ledger height difference from the last snapshot. As the ledger height increases, the snapshot will encompass a larger number of transactions. To understand the impact of ledger height on the rate at which snapshots can be taken,
we conduct tests with different ledger heights: $1000$, $5000$, $10000$, $15000$, and $20000$
as depicted in Fig. \ref{fig:arcthroughput}. We observe that the number of snapshots per minute does not vary much with varying loads, with a median snapshot rate slightly above 100 
snapshots per minute. This trend conveys consistent and constant behavior of the \ours architecture.
Fig. \ref{fig:snapshotpayloadgeneration} represents the relationship between payload size and snapshot generation time. For smaller size payloads (between $0.5$ GB and $2$ GB), the increase in generation time is less. With snapshot payload size increasing beyond $4$ GB, there is a slightly steeper rise in time. Overall, the graph follows a near-linear trend without significant deviation of snapshot time under increasing payload.

\subsection{Comparison with Hyperledger Fabric Snapshot}
As we have extended the existing snapshot mechanism of the Hypelerledger Fabric. It is very important to critically compare the performance of the as-is (HLF offered basic snapshot generation \cite{hlc}) and the enhanced design. For that purpose, we evaluate snapshot generation completion time in both cases. We measure the snapshot creation for various ledger heights and archived transaction contents and capture the response time for the pre-existing HLF snapshot feature as depicted in Fig. ~\ref{fig:snapshotgeneration}. We notice the overall difference between the completion time of the Fabric snapshot (red line) and our combined snapshot-archives implementation with $1K$, $5K$, and $10K$ ledger height is $\approx 0.2 $ seconds. This indicates our archive implementation does not cause any significant overhead. Apart from this, we also captured the bootstrap time when a faulty node is booted from peer storage or from secured encrypted  IPFS storage as depicted in Fig.\ref{fig:snapshotbootstrap}). We observed that, on average, IPFS-based fault node recovery time is only around $1$ second higher than that of local storage, for snapshot archives with $12000$ transactions. 
\begin{figure}[!ht]
\centering
\begin{minipage}{0.48\linewidth}
\centering
\begin{tikzpicture}[scale=0.47]
    \begin{axis}[
        xlabel={\textbf{Ledger Height}},
        ylabel={\textbf{Peer Bootstrap Time(in seconds)}},
        ymin=0, ymax=12,
        xmin=0, xmax=2000,
        xtick={0, 100, 500, 1000, 1500, 2000},
        ytick={1,2,3,4,5,6, 7,8},
        legend pos=north west,
        ymajorgrids=true,
        grid style=dashed,  
    ]

  \addplot[color=brown, mark= diamond]
    coordinates {(100, 5.01)(500, 5.0)(1000, 5.03)(1500, 5.11)(2000, 5.12) }; 
    \addlegendentry{Peer Bootstrap Using Local Store}

    \addplot[color=blue, mark= square]
    coordinates {(100, 5.76)(500, 5.98)(1000, 6.13)(1500, 6.1)(2000, 6.3) }; 
    \addlegendentry{Peer Bootstrap Using IPFS}
    \end{axis}
    \end{tikzpicture}
    \caption{Avg. Peer Bootstrap Time for Recovery}
\label{fig:snapshotbootstrap}
\end{minipage}
\hfill 
\begin{minipage}{0.48\linewidth}
\centering
\begin{tikzpicture}[scale=0.46]
    \begin{axis}[
        xlabel={\textbf{No. of Archived Transactions (X times $1000$)}},
        ylabel={\textbf{Archive Transfer Time Taken (in seconds)}},
        ymin=18, ymax=24.5,
        xmin=0, xmax=32,
        xtick={0, 4, 8, 12, 16, 20, 24, 28, 32},
        ytick={18, 18.5, 19, 19.5, 20, 20.5,21,21.5,22, 22.5},
        legend pos=north west,
        ymajorgrids=true,
        grid style=dashed,  
    ]
    
    \addplot[color=yellow, mark= diamond]
        coordinates {(2, 19.315)(4, 19.276)(8, 19.515)(12, 19.55)(16, 19.519)(20, 19.514)(24, 19.581)(28, 19.819)(32, 19.761)
        };
        \addlegendentry{Ledger Height $1K$}

    \addplot[color=blue, mark= square]
    coordinates {(2, 19.819)(4, 19.746)(8, 19.631)(12, 19.827)(16, 19.932)(20, 19.727)(24, 19.909)(28, 20.172)(32, 20.195)
        };
    \addlegendentry{Ledger Height $5K$}

    \addplot[color=green, mark= triangle]
    coordinates {(2, 20.032)(4, 19.803)(8, 19.791)(12, 19.866)(16, 19.919)(20, 19.856)(24, 19.809)(28, 19.855)(32, 20.27)
        };
    \addlegendentry{Ledger Height $10K$}

    \addplot[color=cyan, mark= star]
    coordinates {(2, 20.119)(4, 20.109)(8, 20.331)(12, 20.117)(16, 20.102)(20, 20.134)(24, 20.306)(28, 20.078)(32, 20.46)
        };
    \addlegendentry{Ledger Height $15K$}
     
    \addplot[color=brown, mark= diamond]
    coordinates {(2, 20.574)(4, 20.651)(8, 21.026)(12, 20.995)(16, 20.874)(20, 21.248)(24, 21.207)(28, 21.144)(32, 21.01)
        };
    \addlegendentry{Ledger Height $20K$}
    
\end{axis}
    \end{tikzpicture}
    \caption{Cross-Network Snapshot Archive Transfer Time}
    \label{fig:snaparc_interop}
\end{minipage}
\end{figure}
\subsection{Snapshot Interoperability}
We also test the performance of our end-to-end interoperable snapshot archive exchange process. This test is also an integral part of our evaluation because when the encrypted snapshot archive data has reached the destination, it has to be cross-verified. So, we have captured the time taken for snapshot archive download, the interoperable framework's response time for transmitting the $\mathtt{CID}$, and sender key information and the archive decryption time at the receiver together to compute the combined transfer time for varied ledger heights with respect to transaction content. We observe that the average response time is a bit higher than the encryption time (Fig.  \ref{fig:snaparc_encr}) or IPFS save time (Fig. \ref{fig:snaparc_up_ipfs}). However, it is expected that this component of the architecture has a lot of interactions in inter-network data exchange. Not only that, but the underlying interoperability framework has to ensure the data is transferred agreeing to the consensus rules and distributed endorsement policies of multiple networks. This component also writes the state updates from the source to the destination ledger immutably. One interesting fact is that the average transfer time does not deviate much and behaves consistently with the increased payload size, as depicted in Fig. \ref{fig:snaparc_interop}. We also notice that after a ledger height of $15K$, the average transfer time slightly increased, but the change does not deviate much with time. 

\subsection{Resilience Test}
\begin{figure*}[hbt!]
\centering
\begin{subfigure}{0.13\linewidth}
\centering
\includegraphics[width=2.4 cm, height=2cm]{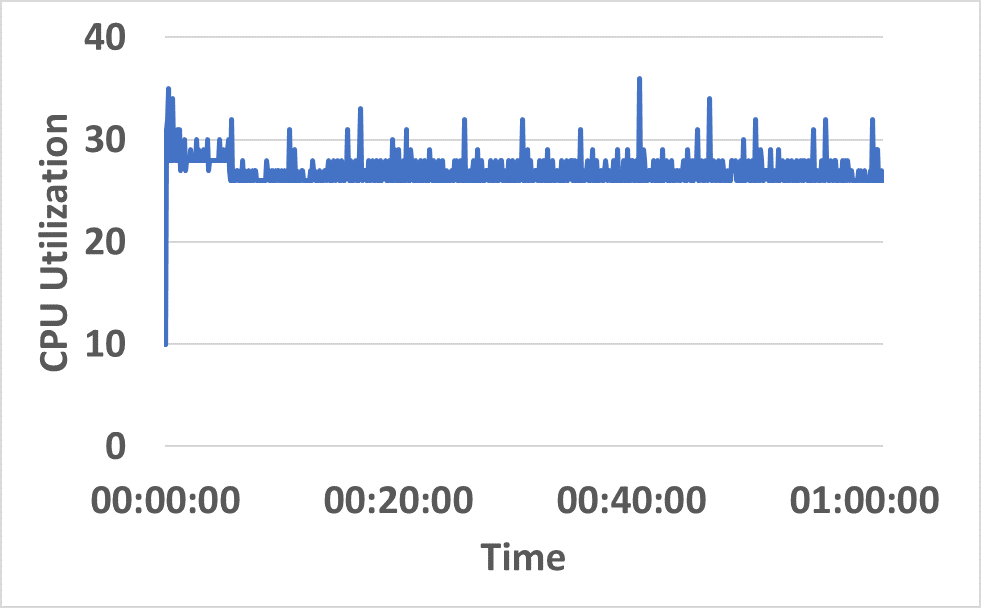}
\caption{\scriptsize $M_{1}$: CPU Usage}
\label{fig:cpu1}
\end{subfigure}
\begin{subfigure}{0.13\linewidth}
\centering
\includegraphics[width=2.4 cm, height=2cm]{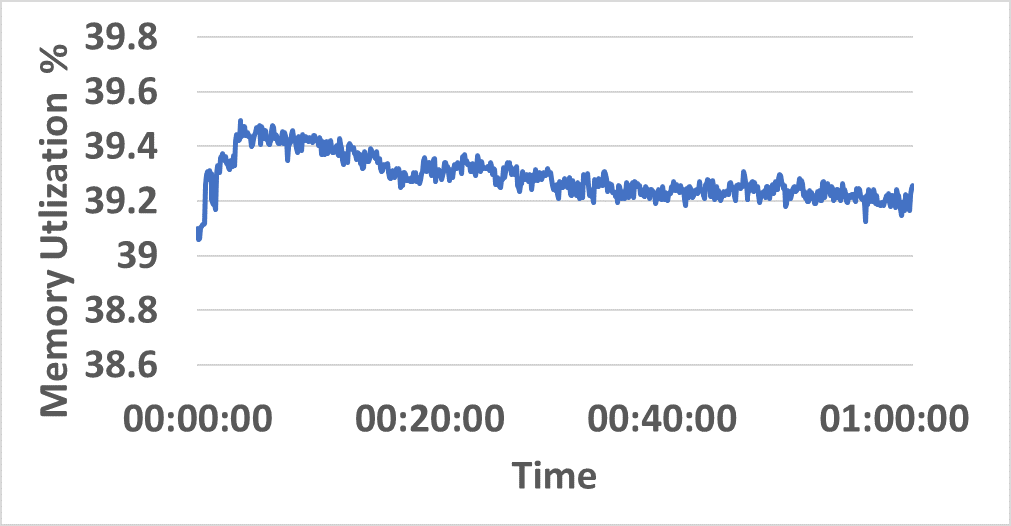}
\caption{\scriptsize $M_{1}$: Mem. Usage}
\label{fig:mem1}
\end{subfigure}
\begin{subfigure}{0.13\linewidth}
\centering
\includegraphics[width=2.45cm,height=2cm]{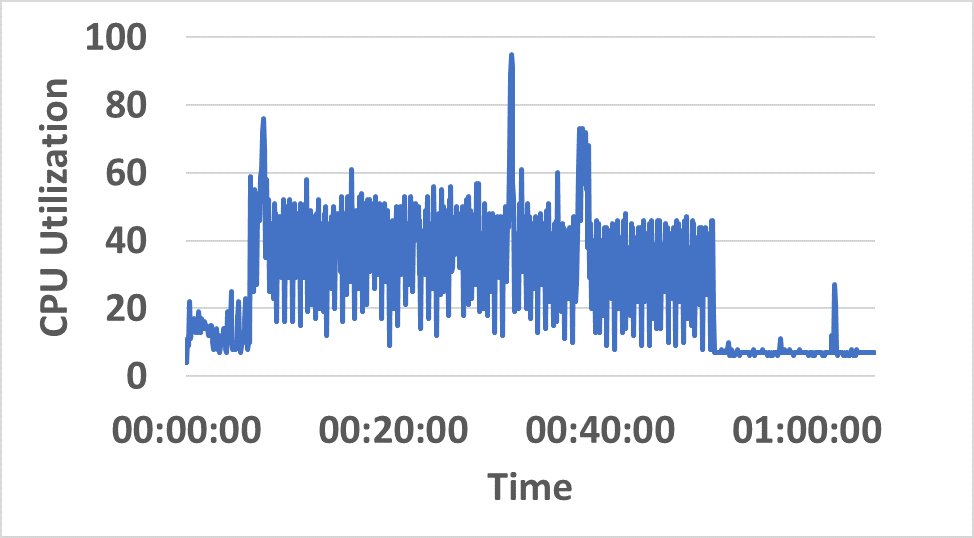}
\caption{\scriptsize $M_{2}$: CPU Usage}
\label{fig:cpu2}
\end{subfigure}
\begin{subfigure}{0.13\linewidth}
\centering
\includegraphics[width=2.45cm, height=2cm]{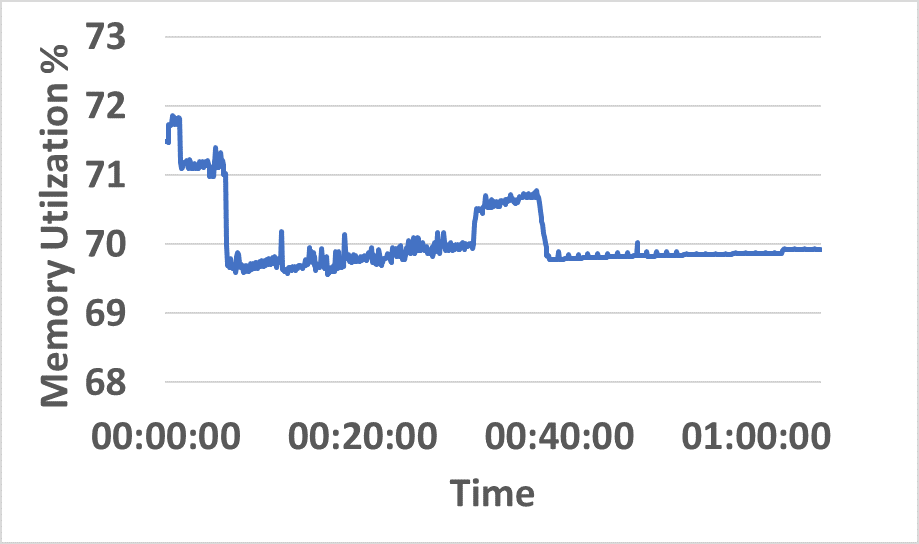}
\caption{\scriptsize $M_{2}$: Mem. Usage}
\label{fig:mem2}
\end{subfigure}
\begin{subfigure}{0.13\linewidth}
\centering
\includegraphics[width=2.4 cm,height=2cm]{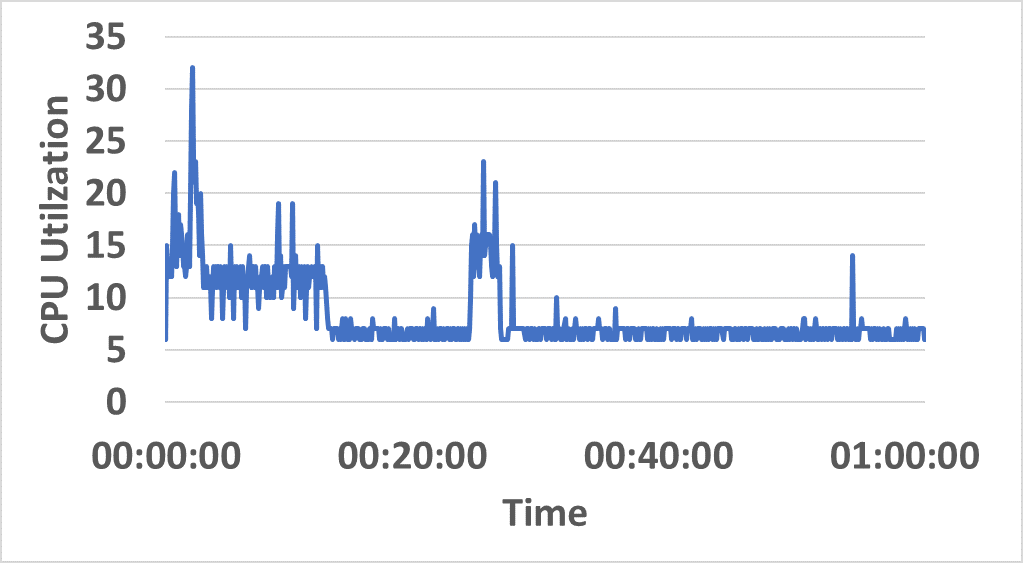}
\caption{\scriptsize $M_{3}$: CPU Usage}
\label{fig:cpu3}
\end{subfigure}
\begin{subfigure}{0.13\textwidth}
\centering
\includegraphics[width=2.45cm, height=2cm]{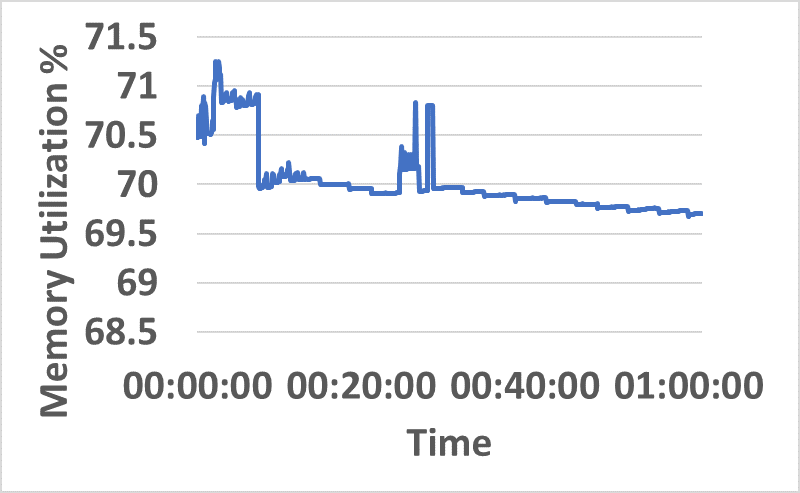}
\caption{\scriptsize $M_{3}$: Mem. Usage}
\label{fig:mem3}
\end{subfigure}
\begin{subfigure}{0.13\textwidth}
\centering
\includegraphics[width=2.45cm, height=2cm]{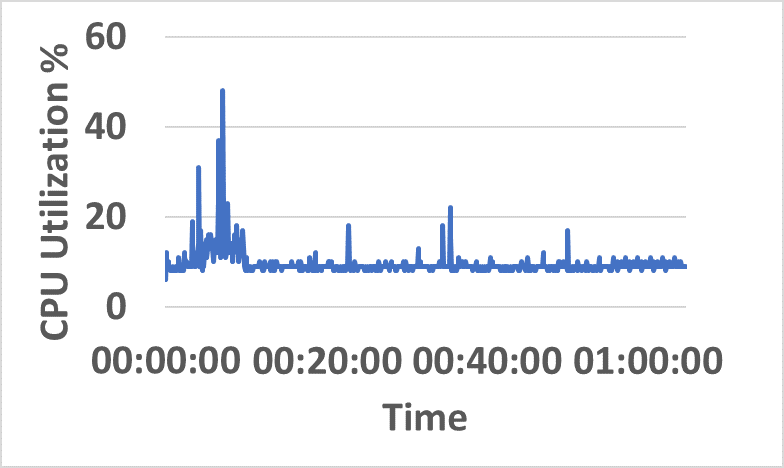}
\caption{\scriptsize $PC$: CPU Usage}
\label{fig:kgpcloudcpu}
\end{subfigure}
\begin{subfigure}{0.13\textwidth}
\centering
\includegraphics[width=2.45cm, height=2 cm]{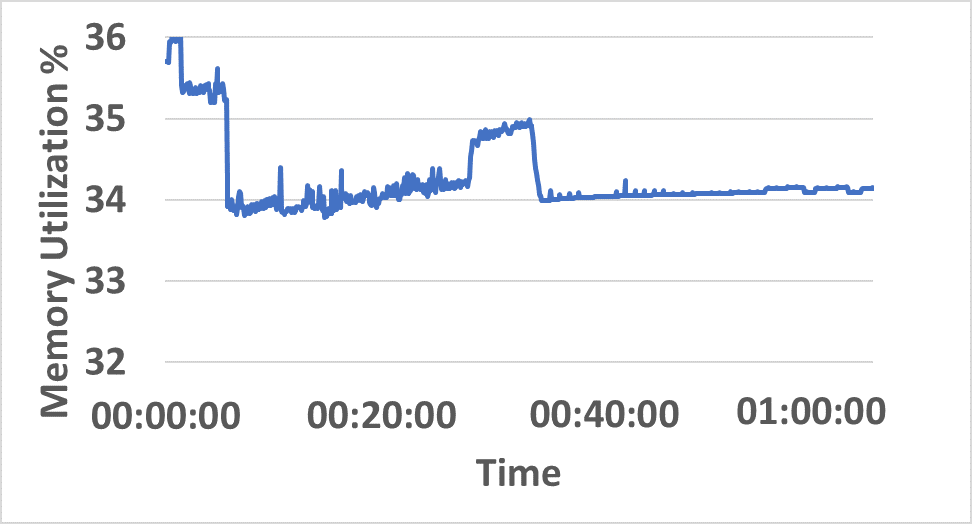}
\caption{\scriptsize $PC$: Mem. Usage}
\label{fig:kgpcloudmem}
\end{subfigure}
\begin{subfigure}{0.13\textwidth}
\centering
\includegraphics[width=2.45cm, height=2 cm]{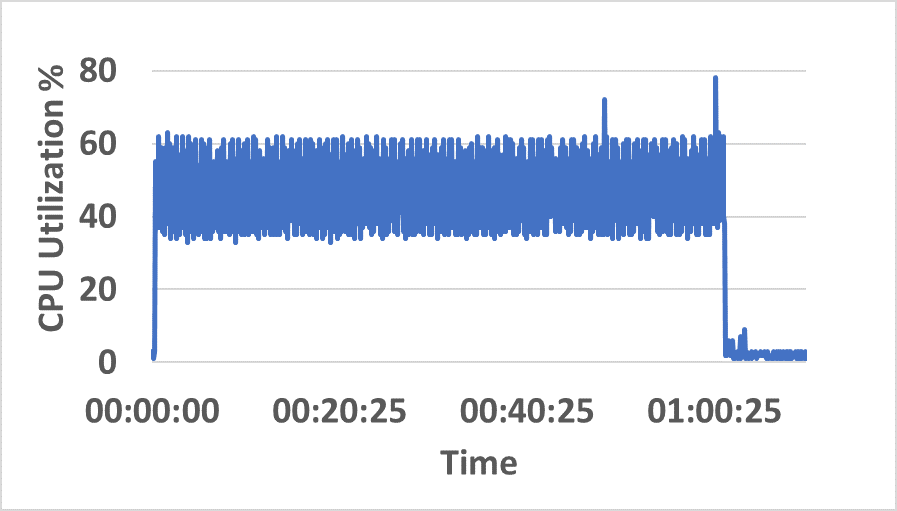}
\caption{\scriptsize $C_{1}$: CPU Usage}
\label{fig:awscpu1}
\end{subfigure}
\begin{subfigure}{0.13\textwidth}
\centering
\includegraphics[width=2.45cm, height=2cm]{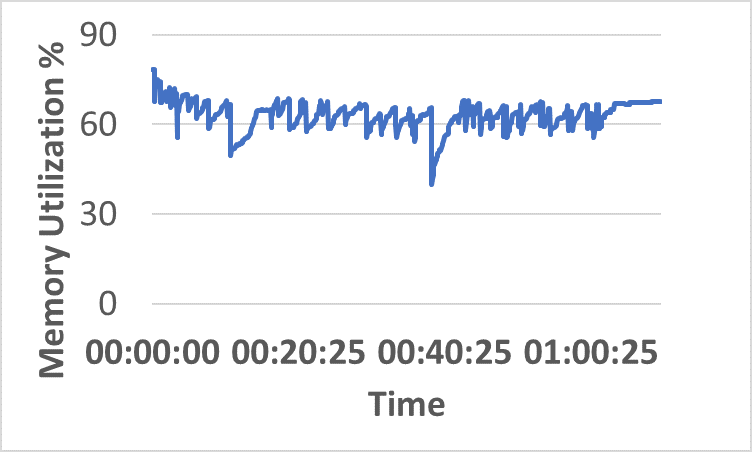}
\caption{\scriptsize $C_{1}$: Mem. Usage}
\label{fig:awsmem1}
\end{subfigure}
\begin{subfigure}{0.13\textwidth}
\centering
\includegraphics[width=2.45cm, height=2 cm]{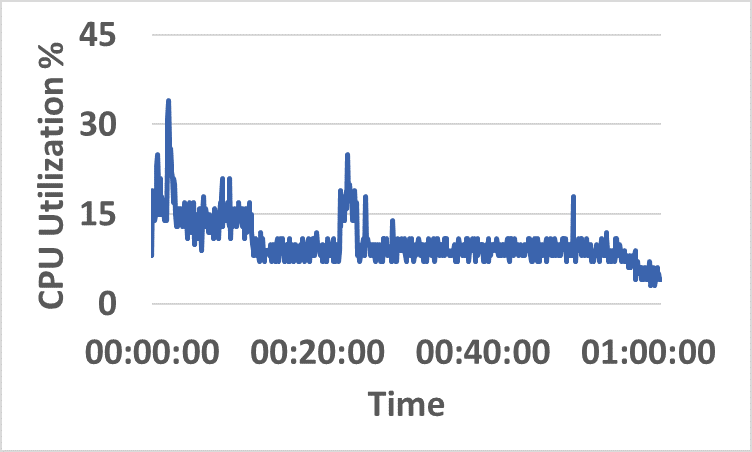}
\caption{\scriptsize $C_{2}$: CPU Usage}
\label{fig:awscpu2}
\end{subfigure}
\centering
\begin{subfigure}{0.13\textwidth}
\centering
\includegraphics[width=2.45cm, height=2cm]{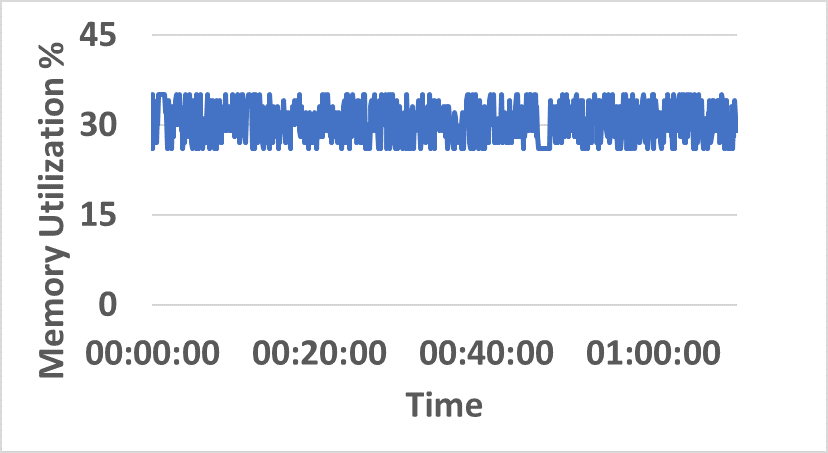}
\caption{\scriptsize $C_{2}$: Mem. Usage}
\label{fig:awsmem2}
\end{subfigure}
\begin{subfigure}{0.13\textwidth}
\centering
\includegraphics[width=2.45cm, height=2cm]{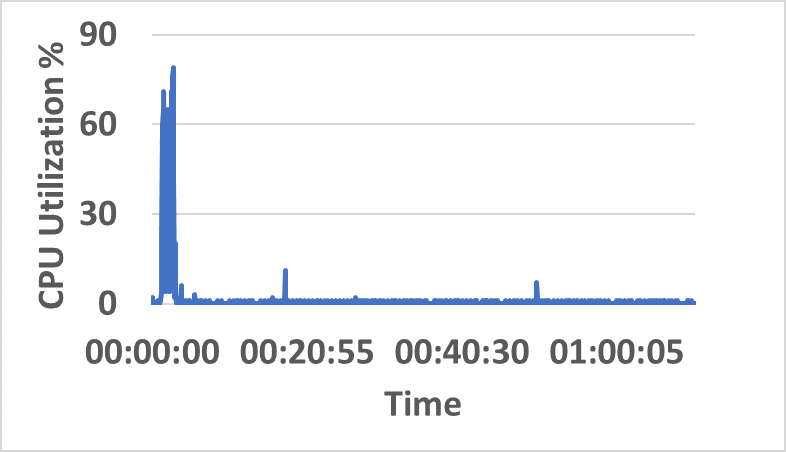}
\caption{\scriptsize $C_{3}$: CPU Usage}
\label{fig:awscpu3}
\end{subfigure}
\begin{subfigure}{0.13\textwidth}
\centering
\includegraphics[width=2.4 cm, height=2 cm]{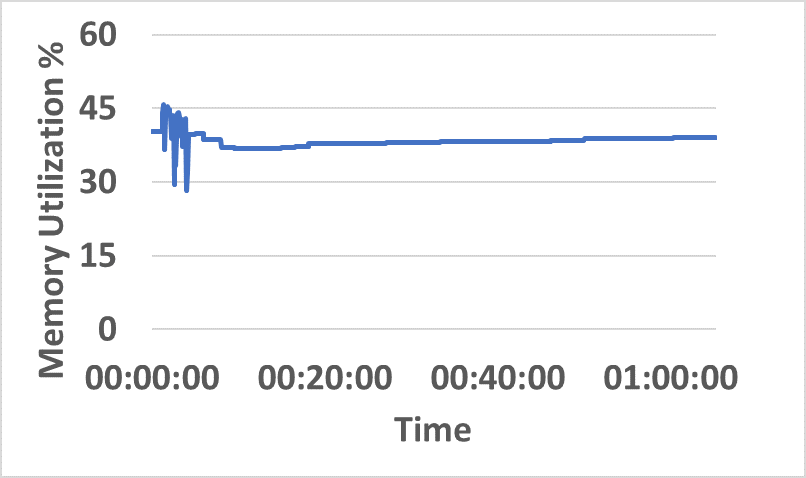}
\caption{\scriptsize $C_{3}$: Mem. Usage}
\label{fig:awsmem3}
\end{subfigure}
\caption{Overall CPU and Memory Usage 
}
\label{fig:cpu_memory_overall}
\end{figure*}

We also run one-hour-long resilience tests for all parts together to observe how the integrated architecture works as a whole. We notice that the system can perform around $900$ transactions without any significant resource strains and performance degradation. Also, an aspect of this test is the success rate of individual transaction components is significantly high ($99.33 \%$) as depicted in Fig.~\ref{fig:transactions}. The number of transactions executed for the interoperability component is slightly lower (less than $5 \%$) than others. This part has a lot of interactions with multiple ledgers and performs cross-chain data communication. Regarding the failures, we found few errors (less than $10$) in the entire test, primarily related to network timeout and data errors. We monitored the on-premises machines, AWS instances, IPFS-hosted machines, and on-premises private cloud infrastructure for resource consumption analysis and the CPU and Memory statistics of each instances as depicted in Fig.~\ref{fig:cpu_memory_overall}. $M_{i}$ in the figure stands for $i^{th}$ laboratory machine. $C_{j}$ refers to the $j^{th}$ AWS cloud machine, and in-house private cloud infrastructure is denoted as $PC$. We observe that the overall CPU utilization ranges from $30 \%$ to $65 \%$, with occasional and momentary increases to $95 \%$. This happened due to some occasional timeout issues under the load. Conversely, we scrutinized the memory usage of all involved resources. Memory utilization ticked between $30 \%$ to $70 \%$ for our hybrid set of resources. Memory usage was also not high, and no resource strain or bottleneck was noted.

\begin{figure}[!hbt]
\centering
\begin{tikzpicture}[scale=0.49] 
\centering
\begin{axis}  
[  
    ybar,  
    enlargelimits=0.15,  
    ylabel={\ \textbf{Average Transactions per minute}}, 
    xlabel={\ \textbf{Type of Transactions}},  
    symbolic x coords={AS,CA,EA,IU,II},
    {ymin=20,ymax=230},
    xtick=data,  
     , 
    nodes near coords align={vertical},
    legend pos=north west,
     ymajorgrids=true,
     grid style=dashed,
    ]  
 \addplot [draw=blue, fill=blue]coordinates {(AS,180)(CA,176) (EA,176)(IU,176)(II,175)};  
\addplot [draw=orange, fill=orange]coordinates {(AS,176) (CA,176) (EA,176) (IU,175) (II,174)};
\addplot [draw=cyan, fill=cyan]coordinates {(AS,4) (CA,0) (EA,0) (IU,1) (II,1)};
\legend{Txns per hour, Pass, Fail}
\end{axis}  
\node[anchor=west, text width=2cm, font=\tiny] at (7, 3) {AS=Archive Snapshot\\ CA=Compress Archive\\ EA=Encrypt Archive\\ IU=IPFS Upload\\ II=Initiate Interop};
\end{tikzpicture}
\caption[Short Title]%
{\centering {\ours Transaction Details}
}
\label{fig:transactions}
\end{figure}

\section{Conclusion and Future Directions}
\label{sec:conclusion_futurework}
\notets{
Unlike existing snapshot and state preserving designs, which primarily focus on generating interoperable blockchain views or state-sharing approaches across multiple networks, \ours introduces an audit-oriented framework that enforces non-repudiation and accountability across permissioned blockchains. 
\ours 
couples cross-chain receipts with snapshot-based archival such that each bilateral exchange is recorded with non-deniable and cryptographically enforced evidence.
Retention of historical snapshots in distributed storage enables post-failure verification and dispute resolution by independent auditors. 
}
Through \ours we not only improve on existing snapshot techniques using need-based snapshot scheduling, snapshot peer selection, and decentralized storage, but also enforce data integrity and privacy during snapshot sharing using content hashing and encryption.
\notets{
In the present scope of this work, the auditor is assumed to act honestly and reliably. However, in real world scenarios the possibility of collusion between an auditor and a participating network can not be entirely ruled out.
To mitigate this limitation, in future works we plan to develop a decentralized and transparent auditing system to compare disputed transactions against snapshots automatically.
}
This system will coordinate the audit process among participating networks, ensuring that all concerned participants are promptly informed about outcomes in a timely and efficient manner. In addition, we also plan to explore incremental snapshotting or selective purging of snapshots to handle large ledger volumes. 
\bibliographystyle{IEEEtran}
\bibliography{sample-base}
    \vspace{-0.5cm}
    \begin{IEEEbiography}[{\includegraphics[width=0.8in,height=1.0in,clip,keepaspectratio]{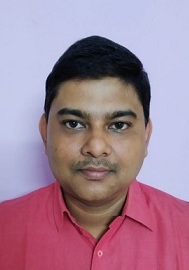}}]{Tirthankar Sengupta}
    (Student Member, IEEE)
    received the M.Tech. degree in Information Technology
    from Indian Institute of Engineering Science and Technology, Shibpur, India. He is at present, a doctoral research
    student in the
    Computer Science and Engineering Department, Indian Institute of Technology, Kharagpur, India. 
    His current research area includes computer security, blockchain,
    and distributed systems.
    \end{IEEEbiography}
    \vspace{-1cm}
   \begin{IEEEbiography}[{\includegraphics[width=0.8in,height=1.0in,clip,keepaspectratio]{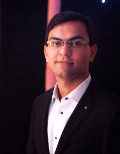}}]{Bishakh Chandra Ghosh}
    received his Ph.D. degree from the Computer Science and Engineering Department at Institute of Technology, Kharagpur, India. He obtained his B.Tech. degree in Information Technology from National Institute of Technology, Durgapur, India. His current research area includes cloud computing, blockchain and distributed systems. 
    \end{IEEEbiography}
\vspace{-1cm}
\begin{IEEEbiography}[{\includegraphics[width=0.8in,height=1.0in,clip,keepaspectratio]{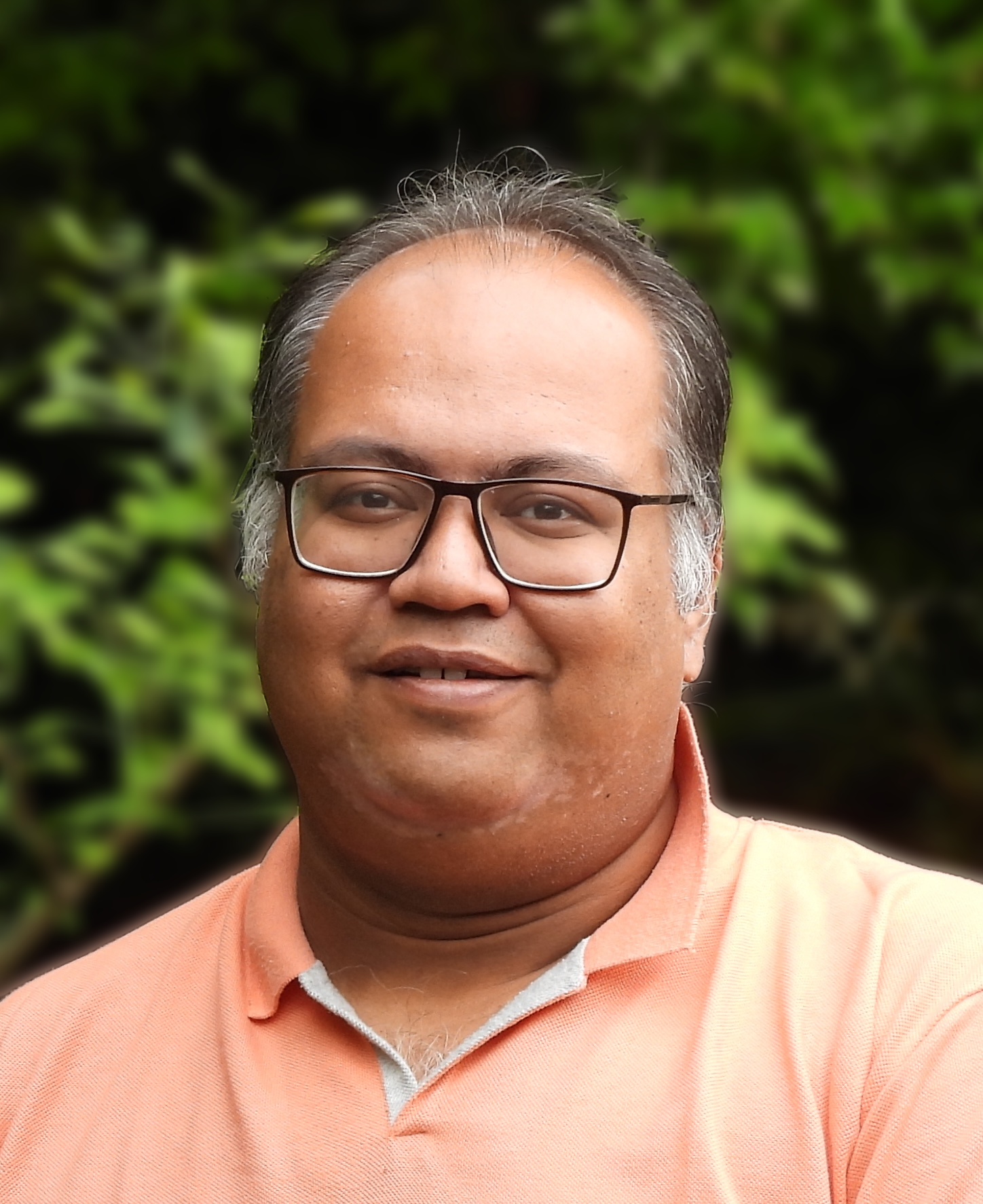}}]{Sandip Chakraborty} (Senior Member, IEEE)
received his Ph.D. degree in Computer Science and Engineering (CSE) from IIT Guwahati, India. Currently, he is an Associate Professor in CSE at IIT Kharagpur, India. He is working as an Area Editors of Elsevier's Ad Hoc Networks, Pervasive and Mobile Computing, and IEEE Transactions on Services Computing. His research interests are on computer systems, pervasive and ubiquitous technologies, and computer-human interactions.
\end{IEEEbiography}
\vspace{-1cm}
\begin{IEEEbiography}
[{\includegraphics[width=0.8in,height=1.0in,clip,keepaspectratio]{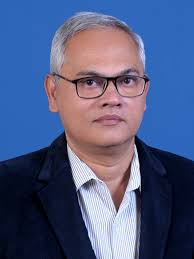}}]{Shamik Sural}(Senior Member, IEEE) 
is a professor in the Department of Computer Science and Engineering at the Indian Institute of Technology, Kharagpur. He is a recipient of Alexander von Humboldt Fellowship
and Fulbright Fellowship. He has served on the editorial
boards of IEEE Transactions on Dependable \&
Secure Computing and IEEE Transactions on
Services Computing. His research interests include computer security, blockchain and data science.
\end{IEEEbiography}
\end{document}